\documentclass[a4paper,twocolumn,11pt,accepted=2023-11-29]{quantumarticle}
\pdfoutput=1
\usepackage[utf8]{inputenc}
\usepackage[english]{babel}
\usepackage[T1]{fontenc}
\usepackage{amsmath}
\usepackage{hyperref}
\usepackage{amssymb}
\usepackage{mathtools}
\usepackage[tight]{subfigure}
\usepackage{enumitem}
\usepackage{soul}
\usepackage{cancel}
\usepackage{multirow}
\usepackage{tikz}
\usepackage{pifont}
\usepackage{xspace}
\usepackage{comment}
\usepackage{physics}
\usepackage{bbold}
\usepackage{graphicx}
\usepackage{dcolumn}
\usepackage{bm}
\usepackage[numbers]{natbib}
\usepackage{ltxgrid}

\hypersetup{colorlinks=true, linkcolor=blue, citecolor=blue, filecolor=blue, urlcolor=blue}

\usepackage[capitalise]{cleveref}

\newcommand{\red}[0]{\color{red}}

\begin{document}
\title{Entanglement dynamics in U(1) symmetric hybrid quantum automaton circuits}

\author{Yiqiu Han}
\email{hankq@bc.edu}
\affiliation{Department of Physics, Boston College, Chestnut Hill, MA 02467, USA}

\author{Xiao Chen}
\affiliation{Department of Physics, Boston College, Chestnut Hill, MA 02467, USA}

\begin{abstract}
    We study the entanglement dynamics of quantum automaton (QA) circuits in the presence of U(1) symmetry. We find that the second R\'enyi entropy grows diffusively with a logarithmic correction as $\sqrt{t\ln{t}}$, saturating the bound established by Ref.\cite{Huang_2020}. Thanks to the special feature of QA circuits, we understand the entanglement dynamics in terms of a classical bit string model. Specifically, we argue that the diffusive dynamics stems from the rare slow modes containing extensively long domains of spin 0s or 1s. Additionally, we investigate the entanglement dynamics of monitored QA circuits by introducing a composite measurement that preserves both the U(1) symmetry and properties of QA circuits. We find that as the measurement rate increases, there is a transition from a volume-law phase where the second R\'enyi entropy persists the diffusive growth (up to a logarithmic correction) to a critical phase where it grows logarithmically in time. This interesting phenomenon distinguishes QA circuits from non-automaton circuits such as U(1)-symmetric Haar random circuits, where a volume-law to an area-law phase transition exists, and any non-zero rate of projective measurements in the volume-law phase leads to a ballistic growth of the R\'enyi entropy.
\end{abstract}

\maketitle

\section{Introduction} 
Entanglement is an important measure of correlations between different degrees of freedom in many-body quantum systems. In a typical system with local interactions, quantum information propagates ballistically, resulting in linear growth of entanglement over time \cite{Kim_2013}. This physics can be understood through random circuit models, which offer a minimal model for investigating entanglement dynamics and information scrambling \cite{Lieb-Robinson,Calabrese_2005,Burrell_2007,Nahum_2017,brown2013scrambling}.

However, the above-described picture changes slightly when an additional continuous symmetry is present in the dynamics. If U(1) symmetry is imposed, it can lead to diffusive transport of the conserved charges. It has been demonstrated that although the von-Neumann entanglement entropy continues to grow linearly, the growth of higher R\'enyi entropies is limited by the diffusive transport and therefore exhibits sub-ballistic growth \cite{Huang_2020,Pollmann_2019,Znidaric_2020,Zhou_2020}. Mathematically it is rigorously proven that the growth of $S^{(n>1)}$ is at most diffusive, with a logarithmic correction \cite{Huang_2020}, i.e., 
\begin{equation}
  S^{(n>1)}\leq \frac{n}{n-1}\mathcal{O}(\sqrt{t\ln t}).
  \label{eq:S_A upper bound}
\end{equation}

Motivated by these findings, this paper investigates the entanglement dynamics in the U(1)-symmetric quantum automaton (QA) circuits with a focus on the second R\'enyi entropy $S^{(n=2)}$. In QA circuits, the quantum state is always an equal-weight superposition of all the allowed basis states with the phases carrying the quantum information. Due to this special property, $S^{(n=2)}$ can be mapped to a quantity of a classical bit string model \cite{QA_Z2,QA_vl,Iaconis_2020}. Such a mapping enables us to study the entanglement dynamics analytically and also provides an efficient method for numerical simulation. We show that the growth of $S^{(n=2)}$ is governed by the presence of the rare bit strings that contain extensively long domains comprising consecutive spin 0s or 1s, consistent with the physical picture introduced in Ref.~\cite{Huang_2020,Pollmann_2019}.  Additionally, we present numerical evidence demonstrating that the dynamics of $S^{(n=2)}$ actually saturates the upper bound defined in Eq.~\eqref{eq:S_A upper bound}. This saturation is caused by the diffusive transport (up to a logarithmic correction) of the boundary of these long domains. Furthermore, for charge-fixed states, we study the coefficient in front of the diffusive scaling of $S^{(n=2)}(t)$ and find that it is linearly dependent on the charge filling factor $\nu$ when $\nu$ is small.

In addition, we are interested in the impact of U(1) symmetry on the entanglement dynamics of monitored QA circuits. Notably, recent research has revealed that monitored quantum dynamics give rise to a measurement-induced entanglement phase transition (MIPT) \cite{Skinner_2019,Chan_2019,Li_2018}. This occurs as a result of the interplay between random unitary evolution and local non-unitary measurements, driving the system  from a highly-entangled volume-law phase to a disentangled area-law phase \cite{Skinner_2019,Chan_2019,Li_2018,Li_2019,Gullans_2020,Bao_2020,Jian_2020}, or even to other quantum phases, such as the critical phase, depending on the symmetry and type of measurements imposed \cite{Chen_2020,Alberton_2021,Ippoliti_2021,Sang_2021,Lavasani_2021,QA_Z2}. When U(1)-symmetry is introduced in monitored Haar random circuits, it is found that any non-zero rate of single-qubit projective measurements will eliminate the rare slow modes containing extensively long domains, and the R\'enyi entropy grows linearly in time for $0<p<p_c$ and exhibits $z=1$ dynamical scaling at the critical point $p_c$ \cite{Charge_sharpening}.

With these insights in mind, our paper also investigates the entanglement dynamics of U(1)-symmetric QA circuits under specific measurements that preserve U(1) symmetry and keep the wave function as an equal weight superposition of basis states.  Interestingly, different from
Haar random circuits, the measurements leave these extensively long domains untouched and the second R\'enyi entropy still exhibits diffusive growth in the volume-law phase. As the measurement rate $p$ increases, we observe a phase
transition to a critical phase where the entanglement entropy grows logarithmically in time. The critical phase that we observe is a result of both the unique properties of QA circuits and the presence of U(1) symmetry. It is worth noting that similar behavior has also been observed in the monitored $\mathbb{Z}_2$ symmetric QA circuits \cite{QA_Z2}.

\section{U(1)-symmetric hybrid QA circuits and two-species particle model}  
In this paper, we consider 1+1d U(1)-symmetric hybrid QA circuits. The dynamics consists of local QA unitary operators and composite measurements, which are chosen to preserve the total charge
\begin{equation}
  Q=\sum_{i}^L \sigma_i, \text{ where }\sigma_i = (1-Z_i)/2
\end{equation}
and $Z_i$ is the Pauli Z matrix acting on the $i$th site of a chain with $L$ qubits. A QA unitary operator permutes states in the computational basis up to a phase, i.e.,
\begin{equation}
  U|n\rangle= e^{i\theta_n}|\pi(n)\rangle,
\end{equation}
where $\pi\in S_N$ is an element of the permutation group on a computational basis with cardinality $N$. Here we take the initial state 
\begin{equation}
    |\psi_0\rangle=\sum_n \frac{|n\rangle}{\sqrt{N}} 
\end{equation}
to be an equal-weight superposition of two different sets of basis states:
(i) $\{|n\rangle=|\sigma_1\sigma_2\dots\sigma_L\rangle: \sigma_i=\{0,1\}\}$ is all the allowed Pauli Z basis with cardinality $N=2^L$ so that 
\begin{equation}
|\psi_0\rangle=|+x\rangle^{\otimes L}=\big[\frac{1}{\sqrt{2}}(|0\rangle+|1\rangle)\big]^{\otimes L}.
\end{equation}
(ii) $\{|n\rangle=|\sigma_1\sigma_2\dots\sigma_L\rangle: \sigma_i=\{0,1\}, \sum_i\sigma_i=Q\}$ is the subset of the Pauli Z basis with a fixed extensive charge filling $\nu\equiv Q/L$, so that $N={L \choose Q}=L!/[(L-Q)!Q!]$.


\begin{figure}[!t]
  \centering
  \includegraphics[width=0.5\textwidth]{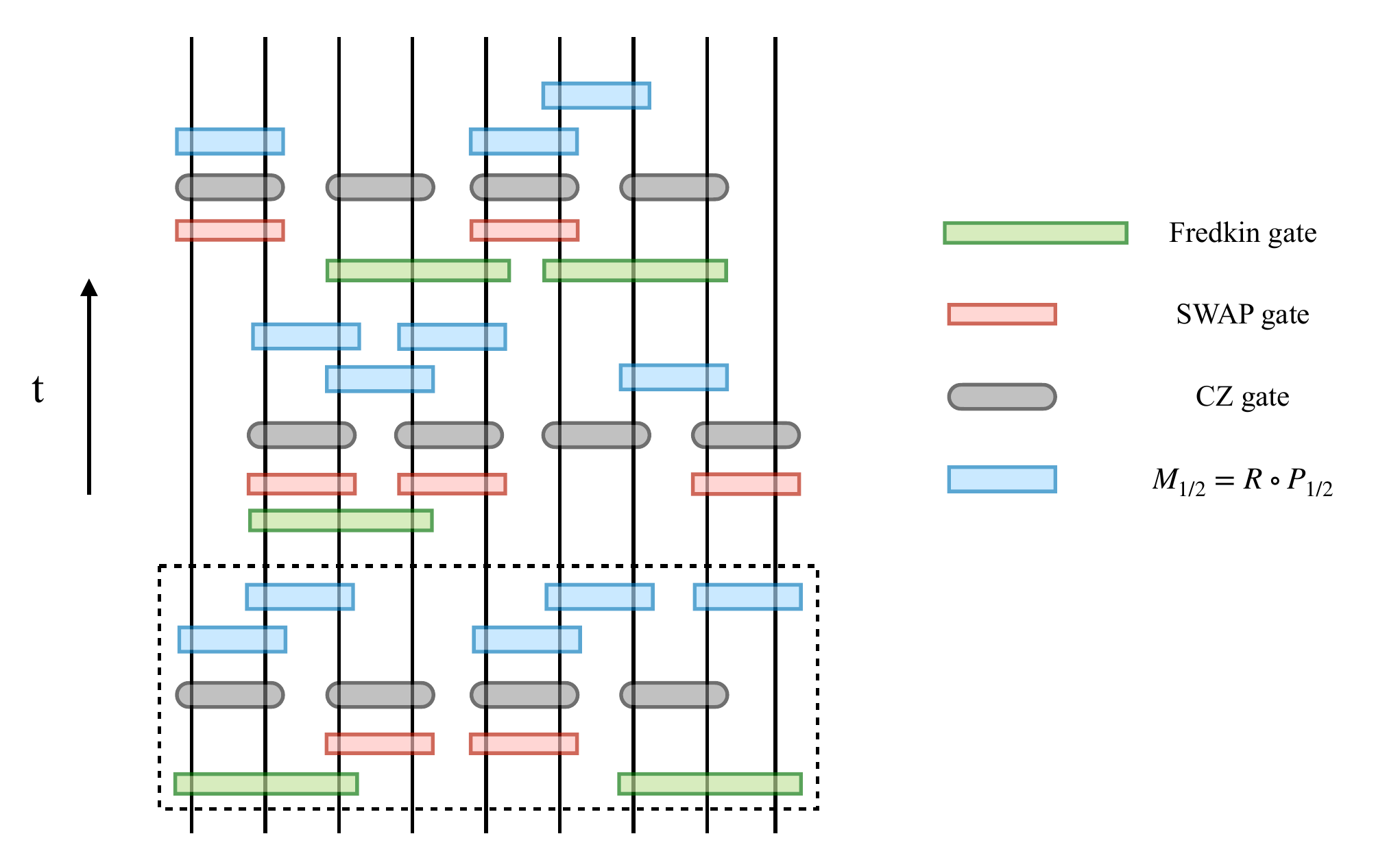}
  \caption{The setup of the U(1)-symmetric hybrid QA circuit of one time step.  The dashed box encloses the gates within a single layer. Each time step involves three layers of Fredkin gates, SWAP gates, and CZ gates, interspersed with composite measurements. The Fredkin and SWAP gates are applied in each layer with probability $p_u<1$, and the measurement appears in each measured sublayer with probability $p$. The Fredkin and SWAP gates determine the dynamics of bit string $|n\rangle$ and can be replaced by other QA gates preserving U(1) symmetry.}
  \label{fig: Fredkin-SWAP setup}
\end{figure}

Fig.\ref{fig: Fredkin-SWAP setup} depicts a brickwork-patterned U(1) symmetric hybrid QA circuit. Each time step consists of three layers of QA unitary operators interspersed with composite measurements with probability $p$. For the unitary part, we consider Fredkin gates and SWAP gates, along with CZ gates which assign a $\pi$ phase to the spin configuration $|11\rangle$. The Fredkin gates are three-qubit gates that interchange qubits $i-1$ and $i+1$ according to the value of the middle (control) qubit, i.e., $|\sigma_{i} 1_{i+1} \sigma_{i+2}\rangle \mapsto |\sigma_{i+2} 1_{i+1} \sigma_{i}\rangle$ and $|\sigma_{i} 0_{i+1} \sigma_{i+2}\rangle \mapsto |\sigma_{i} 0_{i+1} \sigma_{i+2}\rangle$. Meanwhile, the SWAP gates interchange two neighboring qubits. Together with CZ gates, they scramble the quantum information and increase the entanglement entropy of the state until it saturates to the volume-law scaling.  As illustrated in Fig.\ref{fig: Fredkin-SWAP setup}, in the first/ second/ third layer of each time step, the Fredkin gates are applied on sites $\{3j-2,3j-1,3j\}/ \{3j-1,3j,3j+1\}/ \{3j,3j+1,3j+2\}$ for $j\in[1,L/3]$, while the SWAP and CZ gates are applied on sites $\{2j-1,2j\}/ \{2j,2j+1\}/\{2j-1,2j\}$ for $j\in[1,L/2]$. Specifically, we set the occurring probability of the Fredkin and SWAP gates to be $p_u$ and we take $p_u<1$ throughout the paper.

Constructing the measurement gates can be quite tricky. To ensure that $|\psi_t\rangle$ remains an equal-weight superposition of basis states, we introduce a charge-preserving two-qubit composite measurement $M_{1/2}=R\circ P_{1/2}$. Here $P_1$ and $P_2$ are the Kraus operators,
\begin{equation}
    \begin{aligned}
    P_1&=\begin{pmatrix}
        \frac{1}{\sqrt{2}} & 0 & 0 & 0 \\
        0 & 1 & 0 & 0 \\
        0 & 0 & 0 & 0 \\
        0 & 0 & 0 & \frac{1}{\sqrt{2}}
    \end{pmatrix}, \\
    P_2&=\begin{pmatrix}
        \frac{1}{\sqrt{2}} & 0 & 0 & 0 \\
        0 & 0 & 0 & 0 \\
        0 & 0 & 1 & 0 \\
        0 & 0 & 0 & \frac{1}{\sqrt{2}}
    \end{pmatrix},
    \end{aligned}
\end{equation}
followed by a two-site rotation operator,
\begin{equation}
    R=\begin{pmatrix}
  1 & 0 & 0 & 0 \\
  0 & \frac{1}{\sqrt{2}} & \frac{1}{\sqrt{2}} & 0 \\
  0 & \frac{1}{\sqrt{2}} & -\frac{1}{\sqrt{2}} & 0 \\
  0 & 0 & 0 & 1
\end{pmatrix},
\end{equation}
which maps $|01\rangle \mapsto (|01\rangle+|10\rangle)/\sqrt{2}$ and $|10\rangle \mapsto (|01\rangle-|10\rangle)/\sqrt{2}$ \footnote{
In numerical simulations, we omit the extra $\pi$ phase in front of $|10\rangle$ obtained from the rotation of $M_2$. This is because it is equivalent to applying $M_2$ without the extra $\pi$ phase followed by a controlled phase gate. As shown later, such a phase gate will not affect the entanglement phase transition from the volume-law phase to the critical phase. The location of the phase transition is only determined by the corresponding bit string dynamics.}, and acts trivially on $|00\rangle$ and $|11\rangle$, so as to rotate the wave function back to equal weight (up to a phase) superposition of the computational basis. In general, the measurement disentangles the system by discarding the phase information of a quarter of the basis states. However, considering it is a two-qubit operator as required by the U(1) symmetry, as we will see later, the measurements  together with phase gates can actually induce entanglement in certain circumstances. As shown in Fig.\ref{fig: Fredkin-SWAP setup}, each measured layer contains two rows of composite measurements, with each row containing $M_{1/2}$ randomly distributed with probability $p$ on sites $\{2j-1,2j\}/ \{2j,2j+1\}$ for $i\in[1,L/2]$.

Throughout the paper, we focus on the entanglement dynamics between subsystems A and B that the system is bi-partitioned into. Specifically, we consider the second R\'enyi entropy of A, 
\begin{equation}
  \begin{aligned}
    S^{(2)}_A(t) &=-\ln{\text{Tr}[\rho_A^2(t)]},
    \\
    \rho_A(t) &=\text{Tr}_{B} |\psi_t\rangle\langle\psi_t|
  \end{aligned}
\end{equation}
where $|\psi_t\rangle=\tilde{U}_t|\psi_0\rangle$ is the wave function of the quantum trajectory with $\tilde{U}_t$ representing the circuit evolution up to time $t$. We first consider the initial condition (i), where $|\psi_0\rangle=|+x\rangle^{\otimes L}$. In our earlier work, we discovered an efficient algorithm to compute $S_A^{(2)}(t)$ from this initial state \cite{Iaconis_2020}. Additionally, we presented a classical stochastic model to elucidate the entanglement dynamics \cite{QA_Z2}. In the following, we will provide a brief overview of them and apply these methods to our QA circuit with U(1) symmetry.

The purity can be expressed as the expectation of the $\mathsf{SWAP}_A$ operator over double copies of the system \cite{SWAP},
\begin{equation}\label{eq: swap}
  \text{Tr}[\rho_A^2(t)]=\langle\psi_t|_2\otimes\langle\psi_t|_1 \mathsf{SWAP}_A|\psi_t\rangle_1\otimes|\psi_t\rangle_2.
\end{equation}
The $\mathsf{SWAP}_A$ operator swaps the configurations of the copies within region $A$. Since QA circuits preserve the computational basis that spans $|\psi_t\rangle$, we can insert into Eq.\eqref{eq: swap} two sets of complete basis which are acted upon by the circuit in a time-reversed order,
\begin{equation}
  \begin{aligned}
    &\text{Tr}[\rho_A^2(t)]=\sum_{n_1,n_2}\langle\psi_t|_2\langle\psi_t|_1 \mathsf{SWAP}_A|n_1\rangle|n_2\rangle \\
    & \qquad \qquad \qquad \qquad \qquad \quad \langle n_2|\langle n_1|\psi_t\rangle_1|\psi_t\rangle_2
    \\
    &=\frac{1}{4^L} \sum_{n_1,n_2}e^{-i\Theta_{n_1'}(t)}e^{-i\Theta_{n_2'}(t)}e^{i\Theta_{n_1}(t)}e^{i\Theta_{n_2}(t)},
  \end{aligned}
    \label{eq: purity_A}
\end{equation}
where 
\begin{equation}
    e^{i\Theta_{n_i}(t)}=\sqrt{2^L}\langle n_i|\tilde{U}_t|\psi_0\rangle,
    \label{eq:theta_n}
\end{equation}
and
\begin{equation}
  \begin{aligned}
    |n_1'\rangle|n_2'\rangle&\equiv \mathsf{SWAP}_A|n_1\rangle|n_2\rangle
    \\
    &=\mathsf{SWAP}_A|\alpha_1\beta_1\rangle|\alpha_2\beta_2\rangle
    \\
    &=|\alpha_2\beta_1\rangle|\alpha_1\beta_2\rangle,
  \end{aligned}
\end{equation}
where $|\alpha_i\rangle$ and $|\beta_i\rangle$ are the spin configurations in subsystems $A$ and $B$ of $|n_i\rangle$. Therefore, in the numerical simulation, instead of evolving the wave function $|\psi_t\rangle$, we can apply the circuit on the bit strings in a time-reversed order. When evaluating Eq.~\eqref{eq:theta_n} from left to right, although the composite measurement $M$ is non-unitary, we can still derive the effective action on the bit string $\langle n|$. For bit strings that have anti-parallel spins on the sites where $M$ is applied, they are all forced to be either $\langle \dots01\dots|$ or $\langle\dots10\dots|$ after the measurement.

With a few modifications, the above equations used to compute the purity can also be applied to the initial condition (ii). When $|\psi_0\rangle$ is a charge-fixed state of filling factor $\nu$, only the bit string pairs that share the same filling factor both before and after the SWAP, i.e., $\{|n_1\rangle, |n_2\rangle, |n_1^\prime\rangle, |n_2^\prime\rangle\}$ with the same $\nu$, will have nonzero overlap with $|\psi_0\rangle$ and hence contribute to the purity.

\begin{figure}[!t]
  \centering
  \includegraphics[width=0.3\textwidth]{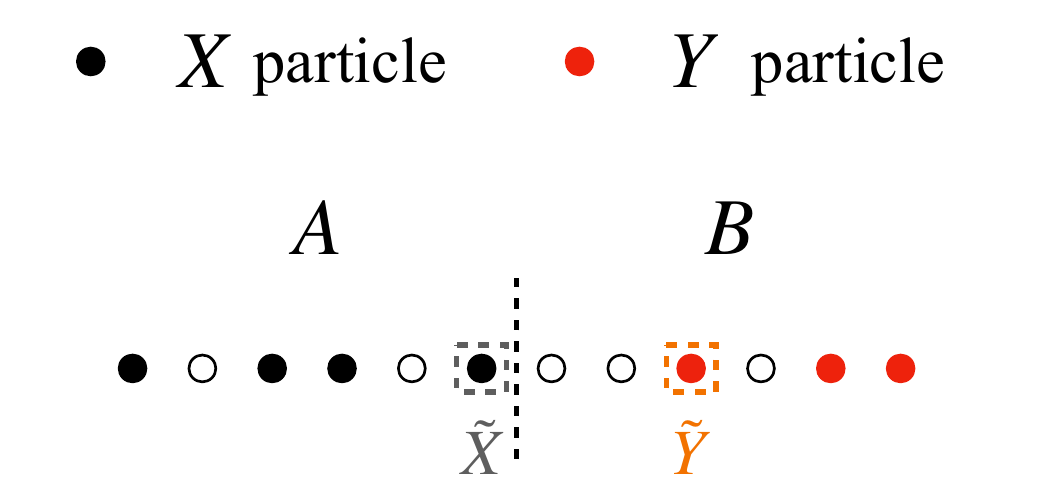}
  \caption{The cartoon of the two-species particle model. We use $\circ$ to denote $h(x)=0$ and $\bullet$ to denote $h(x)=1$, where the black dots represent $X$ particles and the red dots represent $Y$ particles, and $\tilde{X}(\tilde{Y})$ represents the rightmost $X$ (leftmost $Y$) particle. At $t=0$, the $X$ and $Y$ particles are distributed in region $A$ and $B$ respectively. As time evolves, both species begin to expand and intrude into each other's territory. }
  \label{fig:two_species}
\end{figure}

Eq.~\eqref{eq: purity_A} not only offers a numerical method but also  helps us to understand the entanglement dynamics through the classical bit string dynamics. It is worth noting that this equation sums up the accumulated phase $\Theta_r=-\Theta_{n_1'}-\Theta_{n_2'}+\Theta_{n_1}+\Theta_{n_2}$ for each bit string pair $\{|n_1\rangle,|n_2\rangle\}$. Under random time evolution, $\Theta_r$ can become a nonzero random number, and we expect the sum over these phases to be zero. Therefore only the configurations with $\Theta_r=0$ contribute to the purity. This observation leads to a stochastic particle model in which there are two particle species $X$ and $Y$ representing the bit string difference 
\begin{equation}
    h(x,t)=|n_1(x,t)-n_2(x,t)|
\end{equation}
initially distributed in subregion $A$ and $B$ respectively \cite{QA_Z2, QA_vl}, as illustrated in Fig. \ref{fig:two_species}. As time evolves, the two species originally located in subregions $A$ and $B$ gradually expand. It is shown that only the configurations in which $X$ and $Y$ particles have never met up to time $t$ satisfy $\Theta_r(t)=0$ and hence contribute to the purity, i.e.,
\begin{equation}
    S_A^{(2)}(t)= -\ln\text{Tr}\rho_A^2(t)\approx -\ln \frac{N_0(t)}{4^L}\equiv -\ln P(t),
  \label{eq:dyn}
\end{equation}
where $N_0(t)$ is the number of bit string pairs in which the two species never encounter each other up to time $t$. Such an approximation has been numerically verified in Ref.\cite{QA_vl} (For more details, see Appendix.\ref{Appendix: 2ps}).

From the above analysis, the growth of the second R\'enyi entropy is determined by the dynamics of the endpoint $\tilde{X}$ and $\tilde{Y}$ particles of each bit string pair, i.e., the rightmost $X$ particle and the leftmost $Y$ particle. Let us first consider the system without any symmetry. Under unitary dynamics, the $\tilde{X}$ and $\tilde{Y}$ particles move ballistically toward each other at roughly the same speed, i.e., the distance that an endpoint particle travels over time $t$ scales as $\Delta l(t)\propto t$. Therefore, only the configurations whose initial rightmost $X$ and leftmost $Y$ particles are situated a distance of at least $2\Delta l(t)$ apart can contribute to the purity. This leads to $P(t)=[2^{2\Delta l(t)}\times 4^{L-2\Delta l(t)}]/4^L=\mathcal{O}(e^{-t})$, which explains the linear growth of entanglement entropy in the absence of U(1) symmetry.

\section{Unitary dynamics} 
We first study the U(1)-symmetric QA circuit without measurements, i.e., $p=0$, and take the unitary rate $p_u=0.5$. The classical bit string model allows for numerical simulations of the second R\'enyi entropy for relatively large system sizes. To be more specific, we prepare a large sample of randomly generated bit strings which can have either unfixed or fixed charge filling, and estimate $S_A^{(2)}$ using Eq.\eqref{eq: purity_A}. In both cases, we find that the ensemble-averaged early-time $\overline{S^{(2)}_A(t)}$ exhibits a sub-ballistic power-law growth with the exponent close to $1/2$. We also evaluate $P(t)$ by calculating the fraction of the bit string configurations whose corresponding $X$ and $Y$ particles never meet up to time $t$. The numerics indicates that $\overline{-\ln P(t)}$ exhibits the same scaling as the one obtained using Eq.\eqref{eq: purity_A}. Hence we conclude that $-\ln P(t)$ provides a reliable approximation for evaluating $S_A^{(2)}$ in U(1)-symmetric QA circuit. By studying the dynamics of the classical particle model, we can obtain valuable insights into the underlying physics. 

\begin{figure}[!t]
  \centering
  \includegraphics[width=0.45\textwidth]{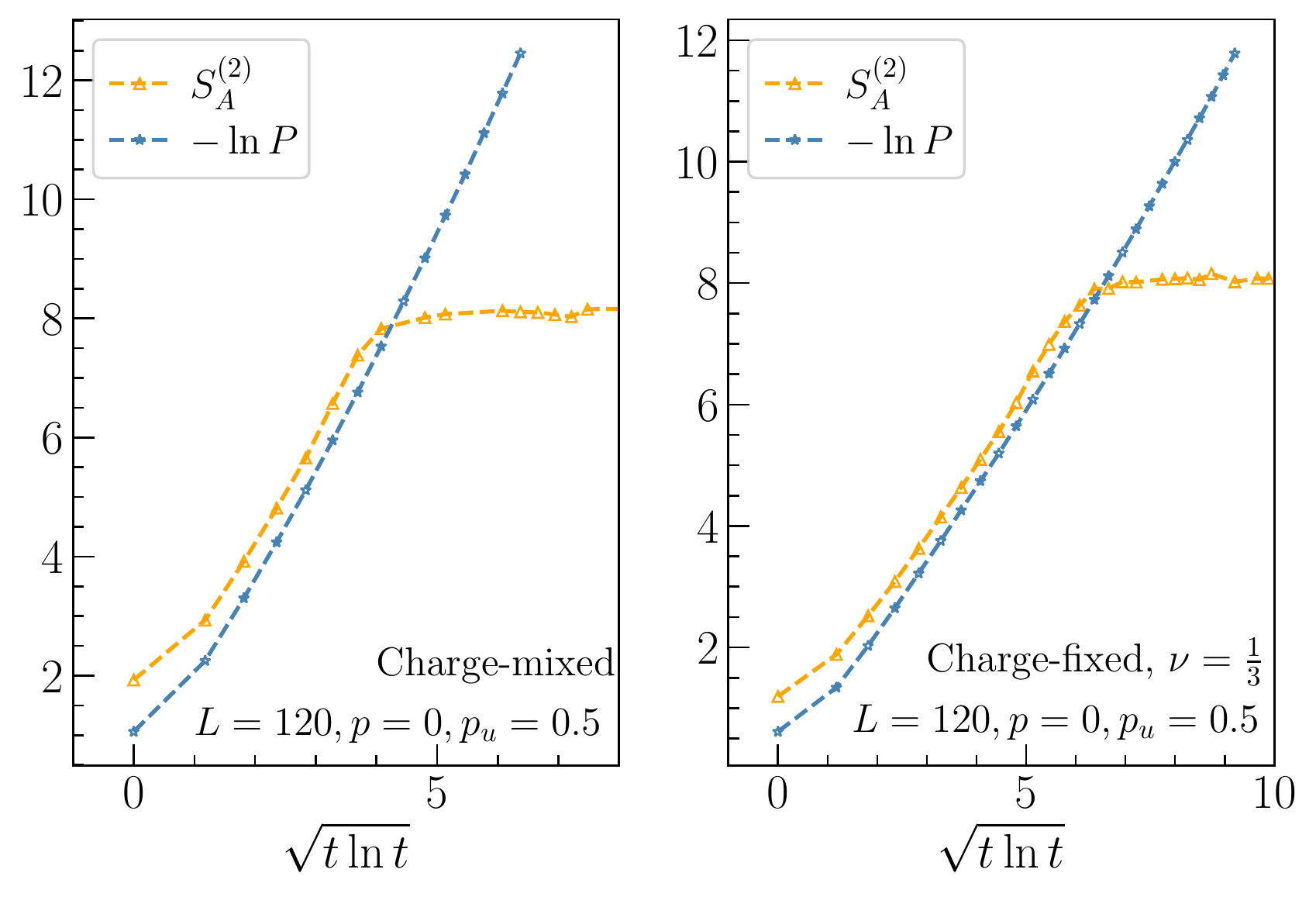}
  \caption{$\overline{S_A^{(2)}(t)}$ and $\overline{-\ln P(t)}$ in the QA circuit with $p=0$. We consider two different initial conditions: the charge-mixed state and the charge-fixed state with the filling factor $\nu=1/3$. We take the unitary rate $p_u=0.5$ in each layer with the system size $L=120$ and subsystem size $L_A=L/2=60$. Both the early-time $\overline{S_A^{(2)}(t)}$ and $\overline{-\ln P(t)}$ for the two initial conditions have $\sqrt{t\ln t}$ scaling.}
  \label{fig:SA_p_0}
\end{figure}

\begin{figure}[!t]
  \centering
  \subfigure[]{\includegraphics[width=0.4\textwidth]{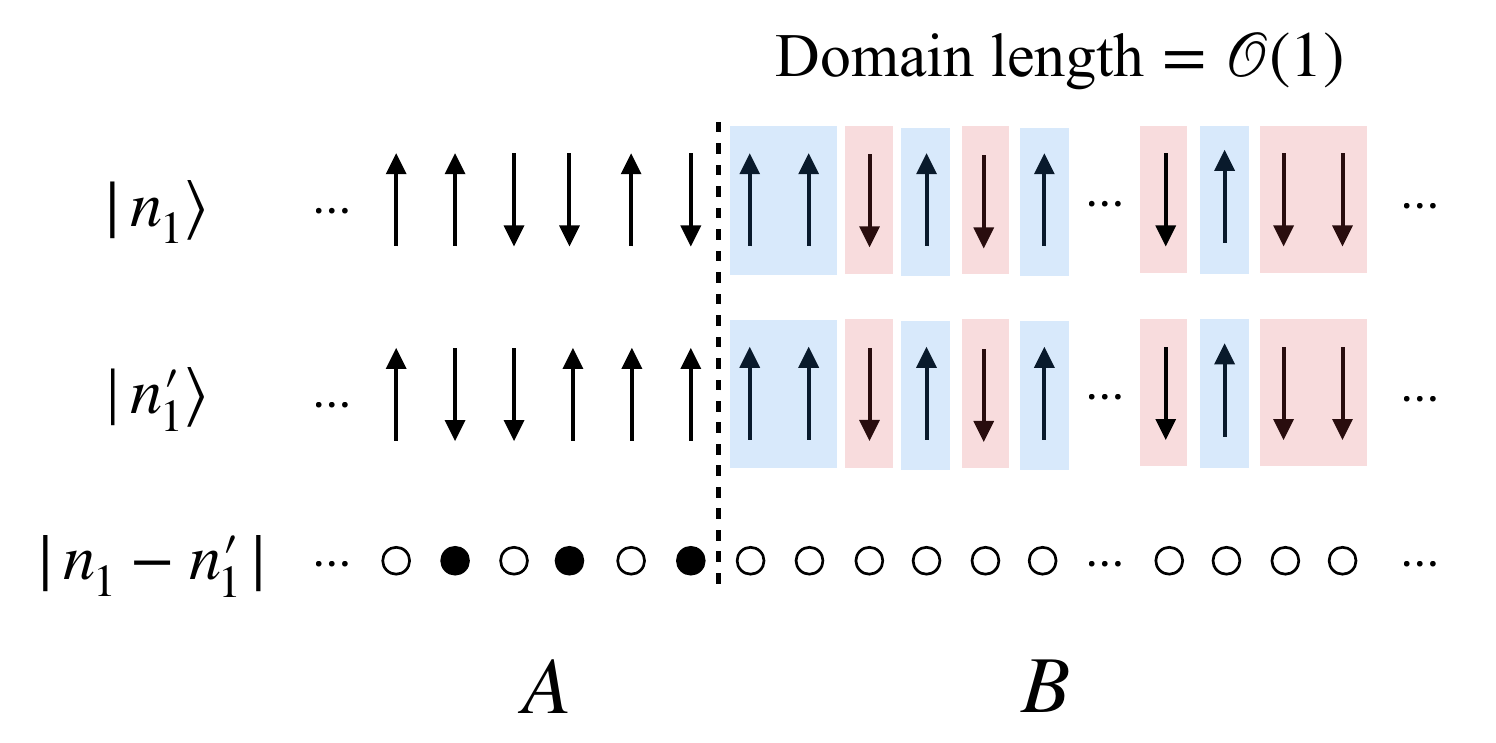}
  \label{fig:fast_mode}}
  \subfigure[]{\includegraphics[width=0.4\textwidth]{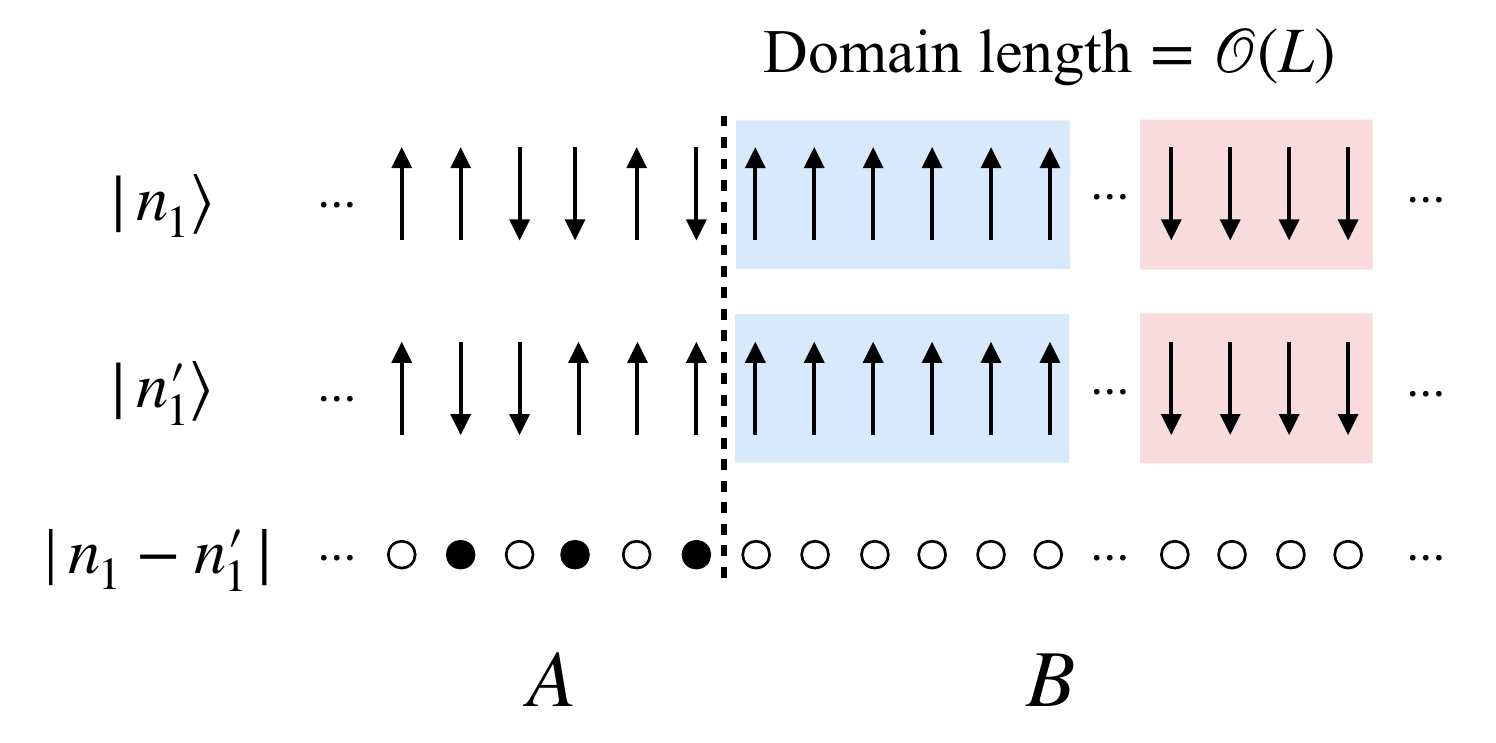}
  \label{fig:slow_mode}}
  \caption{The illustration of the initial bit string configuration of (a) fast modes and (b) slow modes with the same particle representation. For convenience, we consider the bit strings $\{|n_1\rangle, |n_1'\rangle\}$ whose difference represents the particle configuration where there are only $X$ particles located in the region $A$ initially.}
  \label{fig:two_modes}  
\end{figure}

More careful examination of $\overline{-\ln P(t)}$  and $\overline{S_A^{(2)}(t)}$ reveals that the power law growth exponent is slightly larger than $1/2$, which has also been observed in the previous study \cite{Yang_2022}. Since the growth is constrained by Eq.\eqref{eq:S_A upper bound}, the power law exponent cannot exceed $1/2$. We propose that the deviation of the exponent from $1/2$ observed in the numerics is due to the logarithmic correction. As shown in Fig.\ref{fig:SA_p_0}, both quantities are linearly proportional to $\sqrt{t\ln t}$. Therefore, in our QA circuit,
\begin{align}
    \overline{S_A^{(2)}(t)}=\lambda_{EE} \sqrt{t\ln t}.
    \label{eq:EE_logt}
\end{align}

To explain the above results, we study the dynamics of the two-species particle model. Since the dynamics of the $\tilde{X}$ and $\tilde{Y}$ particles are analogous, we will focus on the displacement $\Delta l(t)$ of the $\tilde{X}$ particle. Different from the case without U(1) symmetry where the particles move ballistically, in the presence of U(1) symmetry, there are two distinct modes of $\tilde{X}$ particles depending on the corresponding bit string configuration to the right of $\tilde{X}$ particle. As illustrated in Fig. \ref{fig:fast_mode}, in typical random bit strings whose domains are of $\mathcal{O}(1)$ length, the $\tilde{X}$ particle moves ballistically with a constant velocity. In contrast, in Fig. \ref{fig:slow_mode}, for the bit string configurations with domains of $\mathcal{O}(L)$ length, the $\tilde{X}$ particle only moves diffusively (up to some logarithmic correction). Such configurations are rare and only comprise $\mathcal{O}(e^{-L})$ of the bit string ensemble. To verify the existence of the slow modes, in numerical simulations, we consider the extreme case where the initial bit string configurations in subsystem $B$ are a single domain of spin 0's which is called ``dead region'', i.e.,
\begin{equation}
    \begin{aligned}
    |n_1(t=0)\rangle &=|\alpha_1\rangle \otimes |0\rangle^{\otimes|B|},\\
    |n_1'(t=0)\rangle &=|\alpha_2\rangle \otimes |0\rangle^{\otimes |B|}.
    \end{aligned}
\end{equation}
We also study $\Delta l(t)$ without the dead region for comparison.

\begin{figure}[!t]
  \centering
  \subfigure[]{\includegraphics[width=0.235\textwidth]{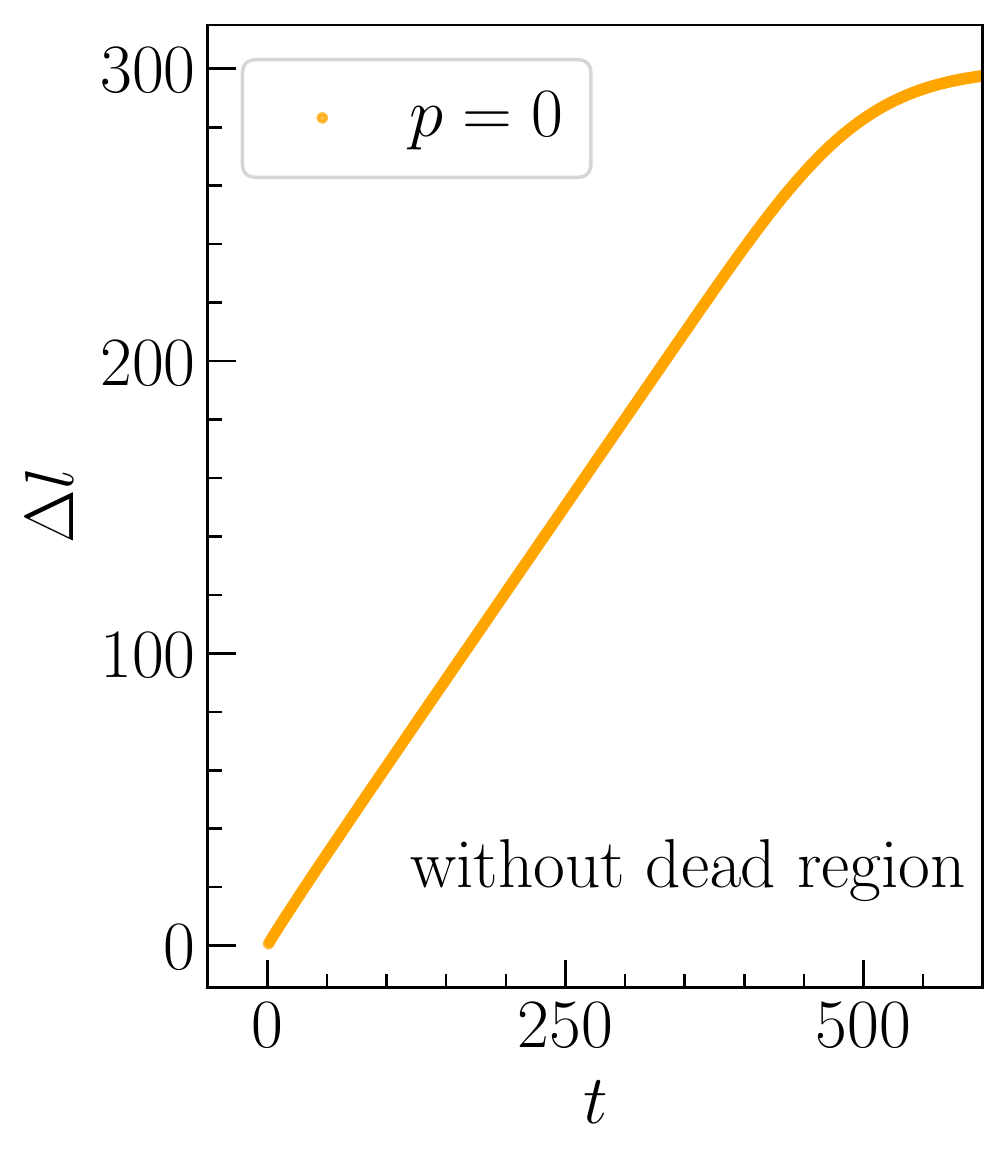}}
  \subfigure[]{\includegraphics[width=0.235\textwidth]{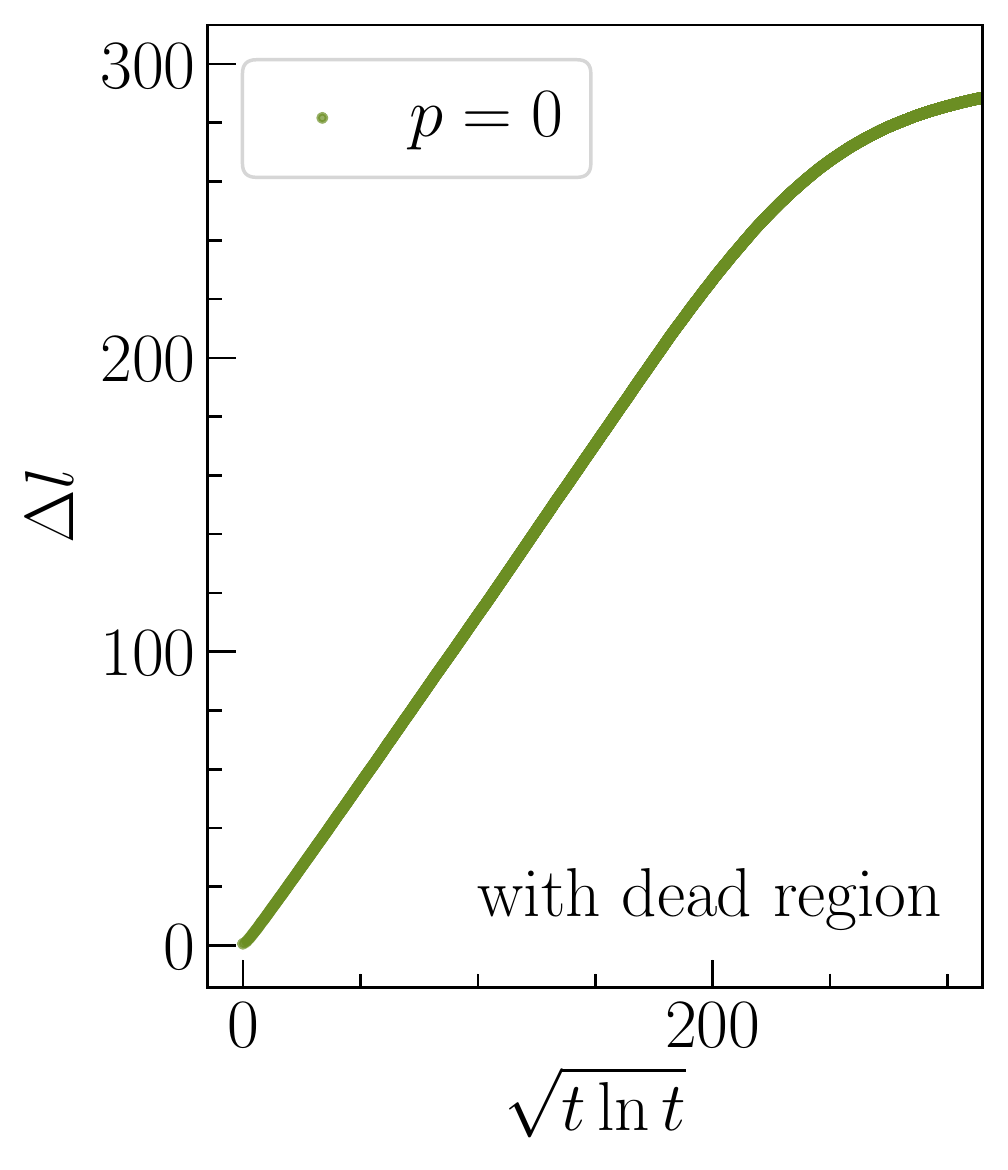}}
  \caption{The distance that an endpoint particle travels $\Delta l(t)$ over time $t$ under U(1)-symmetric QA unitaries for the initial configurations (a) without the dead region and (b) with the dead region. Without loss of generality, we take the probability of Fredkin and SWAP gates to be $p_u=0.5$ in each layer, the system size $L=600$, and the number of spin 1's to be $L/3$ in (a) and $L_A/3$ in (b), both of which have $\nu=1/3$. We find that (a) $\overline{\Delta l}\propto t$ and (b) $\overline{\Delta l}\propto \sqrt{t\ln t}$ before saturation. }
  \label{fig:delta_l}  
\end{figure}

As shown in Fig. \ref{fig:delta_l}, for the configurations without the dead region, $\overline{\Delta l}$ grows linearly in time. This is responsible for the ballistic information spreading observed in the out-of-time-ordered correlator (OTOC) in a similar U(1) symmetric QA circuit \cite{Chen_2020}. On the other hand, for configurations with the dead region, $\overline{\Delta l}$ exhibits diffusive growth over time, with a logarithmic correction, that is, 
\begin{equation}\label{eq: l_logt}
    \overline{\Delta l}=\lambda_l\sqrt{t\ln t}. 
\end{equation}
The diffusive motion of the $\tilde{X}$ particle comes from the diffusive dynamics of the rightmost charge (spin 1) located at the boundary of the dead region. In the simple symmetric exclusion process, one of the most basic models with U(1) symmetry, it is analytically proven that the displacement of the rightmost charge expands as $\sqrt{t\ln{t}}$ \cite{symmetric_exclusion}. Despite the greater complexity of our model involving the Fredkin gate, we believe that the underlying physics remains fundamentally the same. In Appendix.\ref{Appendix: other models}, we examine a simple QA circuit with the kinetic constraint set by SWAP gates only, enabling an exact mapping of the bit string dynamics to the simple symmetric exclusion processes. In addition, we also investigate another QA circuit involving a four-qubit gate. For both models, we provide numerical evidence confirming the existence of logarithmic corrections in both Eq.\eqref{eq: l_logt} and Eq.\eqref{eq:EE_logt}.



Based on this analysis, the bit string pairs that contribute to the purity (See Eq.\eqref{eq:dyn}) can be divided into two parts, $P(t)=P^F(t)+P^S(t)$, where $P^F(t)$ and $P^S(t)$ are the fractions of fast modes and slow modes respectively whose $X$ and $Y$ particles have \textit{not} encountered each other up to time $t$. Since the distance between the endpoint $X$ and $Y$ particles decreases by $2\Delta l(t)$ over time $t$, only the configurations whose initial two species are located a distance $2\Delta l(t)$ apart contribute to $P(t)$. Therefore, both $P^F(t)$ and $P^S(t)$ decay as $\exp(-\overline{\Delta l(t)})$, whereas $P^F(t)\propto \exp(-t)$ and $P^S(t)\propto\exp(-\sqrt{t\ln{t}})$. In the absence of U(1) symmetry, the bit string ensemble comprises only fast modes, which account for the linear growth of $S_A^{(2)}(t)$ as explained earlier. In the presence of U(1) symmetry, slow modes consisting of the bit string pairs with identical long domains with length $\mathcal{O}(L)$ between $\tilde{X}$ and $\tilde{Y}$ emerge. Therefore it takes $\mathcal{O}(L^2)$ time for $\tilde{X}$ particle to reach $\tilde{Y}$ particle. For $S_A^{(2)}(t)\approx -\ln[P^F(t)+P^S(t)]$, $P^F(t)$ vanishes at time $\mathcal{O}(L)$, leaving a diffusive growth of $S_A^{(2)}(t)$ caused by the slow modes up to time $\mathcal{O}(L^2)$.

\begin{figure}[!t]
  \centering
  \includegraphics[width=0.4\textwidth]{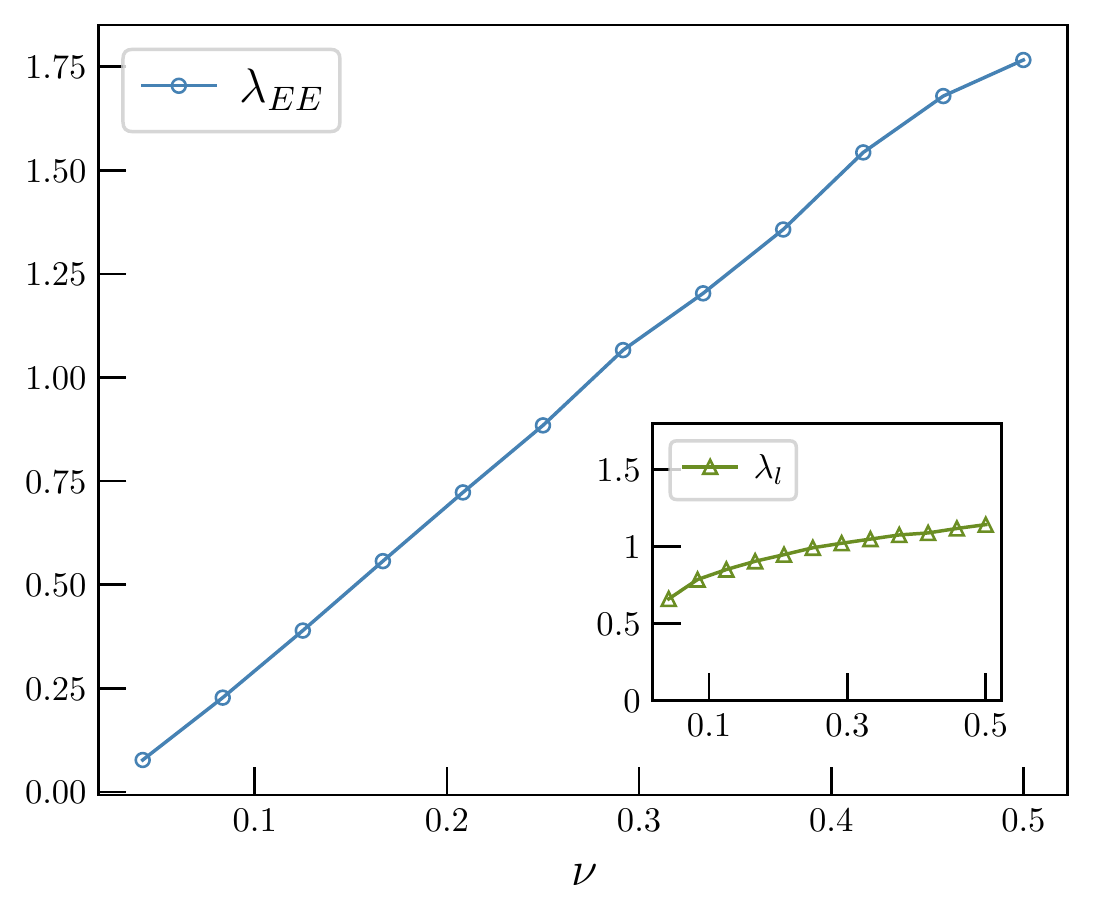}
  \caption{The coefficient $\lambda_{EE}$ of $\overline{S_A^{(2)}}=\lambda_{EE}\sqrt{t\ln t}$ vs the filling factor $\nu$. In the inset, we present $\lambda_l$ of the endpoint displacement $\overline{\Delta l}=\lambda_l\sqrt{t\ln t}$ of the slow modes with dead region vs the filling factor $\nu_A$ in subsystem $A$.}
  \label{fig:SA_lambda}
\end{figure}

Similar reasoning can also be applied to the charge-fixed state. In particular, we can also analyze the dependence of the coefficient $\lambda_{EE}$ in Eq.\eqref{eq:EE_logt} on the filling factor $\nu$. As shown in Fig.~\ref{fig:SA_lambda}, $\lambda_{EE}\propto\nu$ for $\nu\lessapprox 0.3$. To understand this behavior, we  investigate the dependence of $P^S$ on $\nu$, given by
\begin{widetext}
\begin{equation}
    P^S=\left[{L-2\Delta l \choose \nu L}\bigg/{L \choose \nu L}\right]^2 =\left[\frac{(L-2\Delta l)!(L-\nu L)!}{(L-2\Delta l-\nu L)!L!}\right]^2.
\end{equation}
\end{widetext}
If $\nu\ll 1-2\Delta l/L$, we can approximate as follows:
\begin{widetext}
\begin{equation}
    \begin{aligned}
        -\ln{P^S}&\approx -2[(L-2\Delta l)\ln{(L-2\Delta l)}+(L-\nu L)\ln{(L-\nu L)}-(L-2\Delta l-\nu L)\ln{(L-2\Delta l-\nu L)}-L\ln{L}] \\ 
        &=-4\Delta l\ln{\left(1-\frac{\nu L}{L-2\Delta l}\right)}-2\nu L\ln{\left(1-\frac{2\Delta l}{L-\nu L}\right)}+\mathcal{O}(\ln{\Delta l})+\mathcal{O}(\ln{\nu L}) \\
        &\approx 8\nu\Delta l+\mathcal{O}(\ln{\Delta l})+\mathcal{O}(\ln{\nu L}).
    \end{aligned}
\end{equation}
\end{widetext}
As time evolves, the entanglement entropy is dominated by the slow modes and has the scaling $S_A^{(2)}(t)\approx -\ln{P^S(t)}\propto \nu\Delta l$. 
 We further investigate the dependence of $\Delta l$ on $\nu$. We examine the displacement of $\tilde{X}$ particle of the bit string configurations with the dead region as illustrated in Fig.~\ref{fig:slow_mode}. Specifically, we define $\nu_A$ to be the filling factor in subsystem $A$. As shown in Fig.~\ref{fig:SA_lambda}, $\Delta l$ remains largely unaffected by changes in $\nu_A$. 


Combining these findings, we can explain the numerical observation of the linear dependence of $\lambda_{EE}$ on $\nu$ when $\nu$ is small. In addition, we also observe similar linear dependence on $1-\nu$ for $\nu\gtrapprox 0.7$ (not presented in the figure), which can be understood in a similar way.

\section{Hybrid dynamics} 
\begin{figure}[!t]
  \centering
  \subfigure[]{\includegraphics[width=0.4\textwidth]{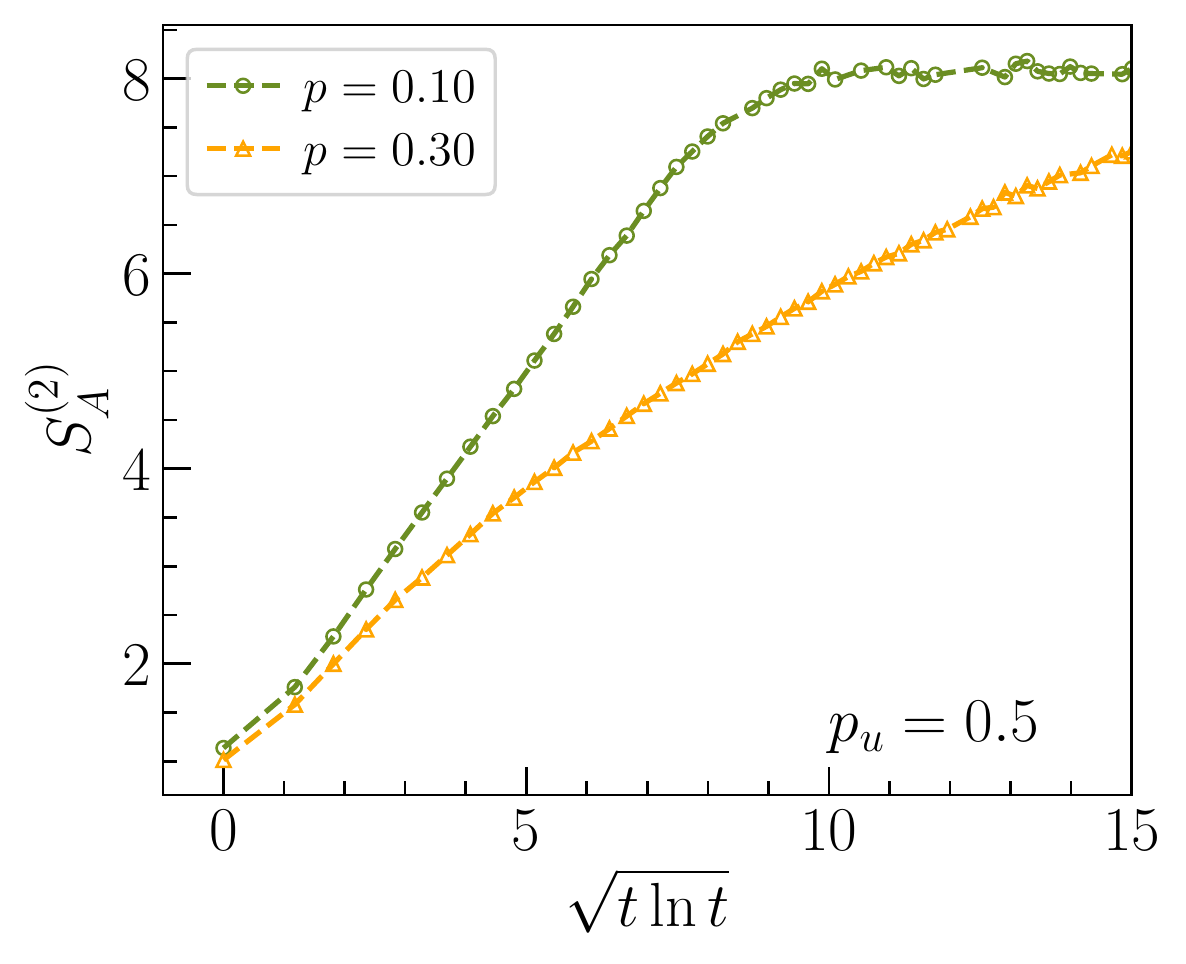}}
  \subfigure[]{\includegraphics[width=0.4\textwidth]{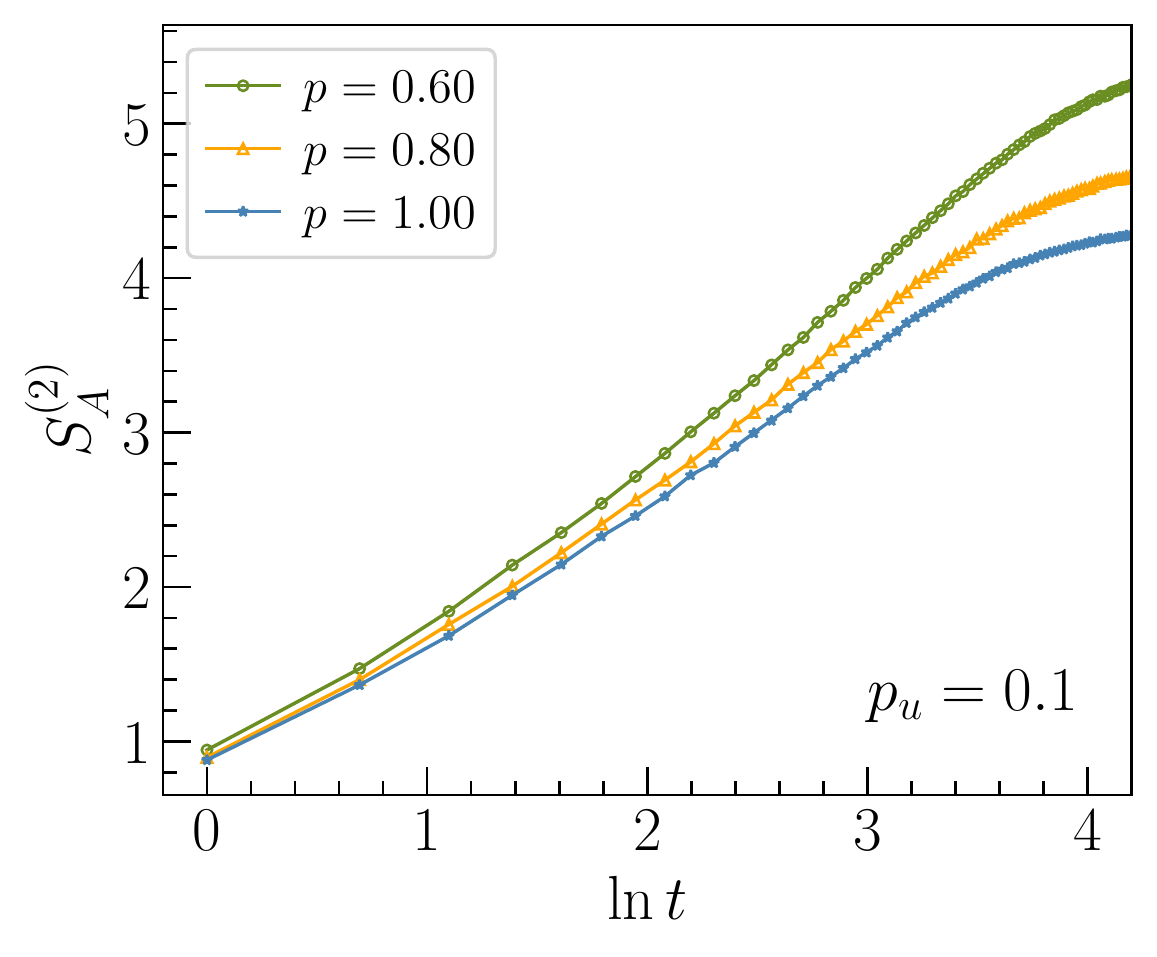}}
  \caption{(a) $S_A^{(2)}$ vs $\sqrt{t\ln{t}}$ for $p=0.1$ and $p=0.3$ with the unitary rate $p_u=0.5$. (b) $S_A^{(2)}$ vs $\ln{t}$ for $p=0.6$, 0.8 and 1 with $p_u=0.1$. For both phases, we take the system size $L=120$ and the filling factor $\nu=1/3$, under the periodic boundary condition (PBC). }
  \label{fig:SA_p>0}
\end{figure}

Now we introduce the composite measurements into the circuit and examine its impact on the entanglement dynamics. As shown in Fig.\ref{fig:SA_p>0}, our finding indicates that when $p$ is small, the entanglement still grows diffusively with a logarithmic correction, as described in Eq.\eqref{eq:EE_logt}.  This can be explained by the fact that our composite measurement only acts non-trivially on anti-parallel neighboring sites, while leaving the bit strings with extensively long domains unaffected. We expect that this scaling behavior persists throughout the entire volume-law phase. This is different from the single-qubit projective measurement which quickly destroys the slow modes with dead regions and leads to the linear growth of $S_A^{(2)}$ in the volume-law phase of the non-unitary U(1) symmetric Haar random circuit\cite{Charge_sharpening}.

As $p$ increases, we observe a decrease in the coefficient $\lambda$ for $S_A^{(2)}(t)$. Eventually, this diffusive growth is replaced by logarithmic growth. To expand the tuning range for the ratio $p/p_u$, we fix $p$ to be a finite constant and reduce $p_u$. Surprisingly, we find that even when $p_u$ approaches zero, the logarithmic scaling persists. This observation suggests that there is an entanglement phase transition from a volume-law phase to a critical phase in our model.

\begin{figure*}[!t]
  \centering
  \subfigure[]{\includegraphics[width=0.4\textwidth]{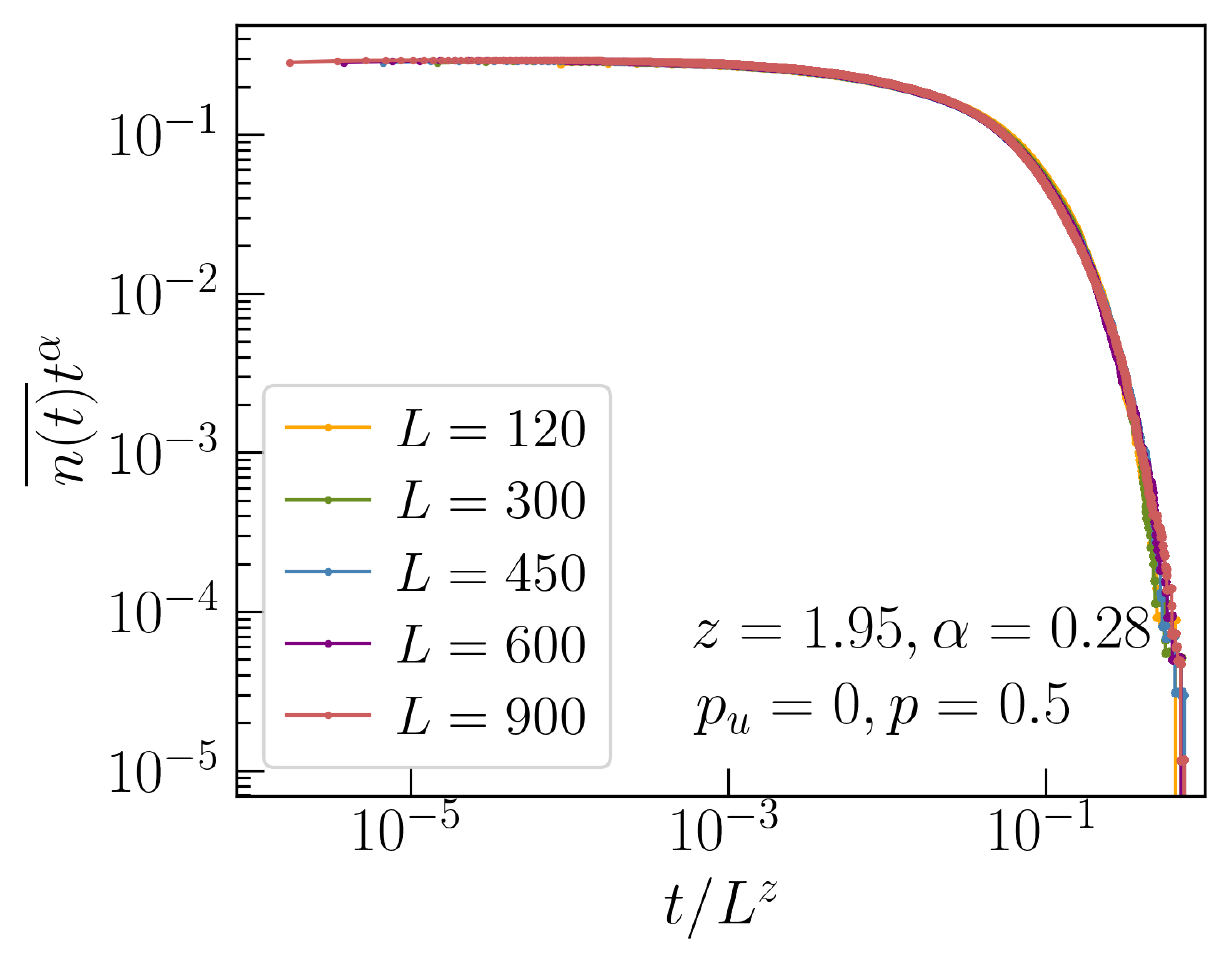}\label{fig:p_u=0}}%
  \subfigure[]{\includegraphics[width=0.4\textwidth]{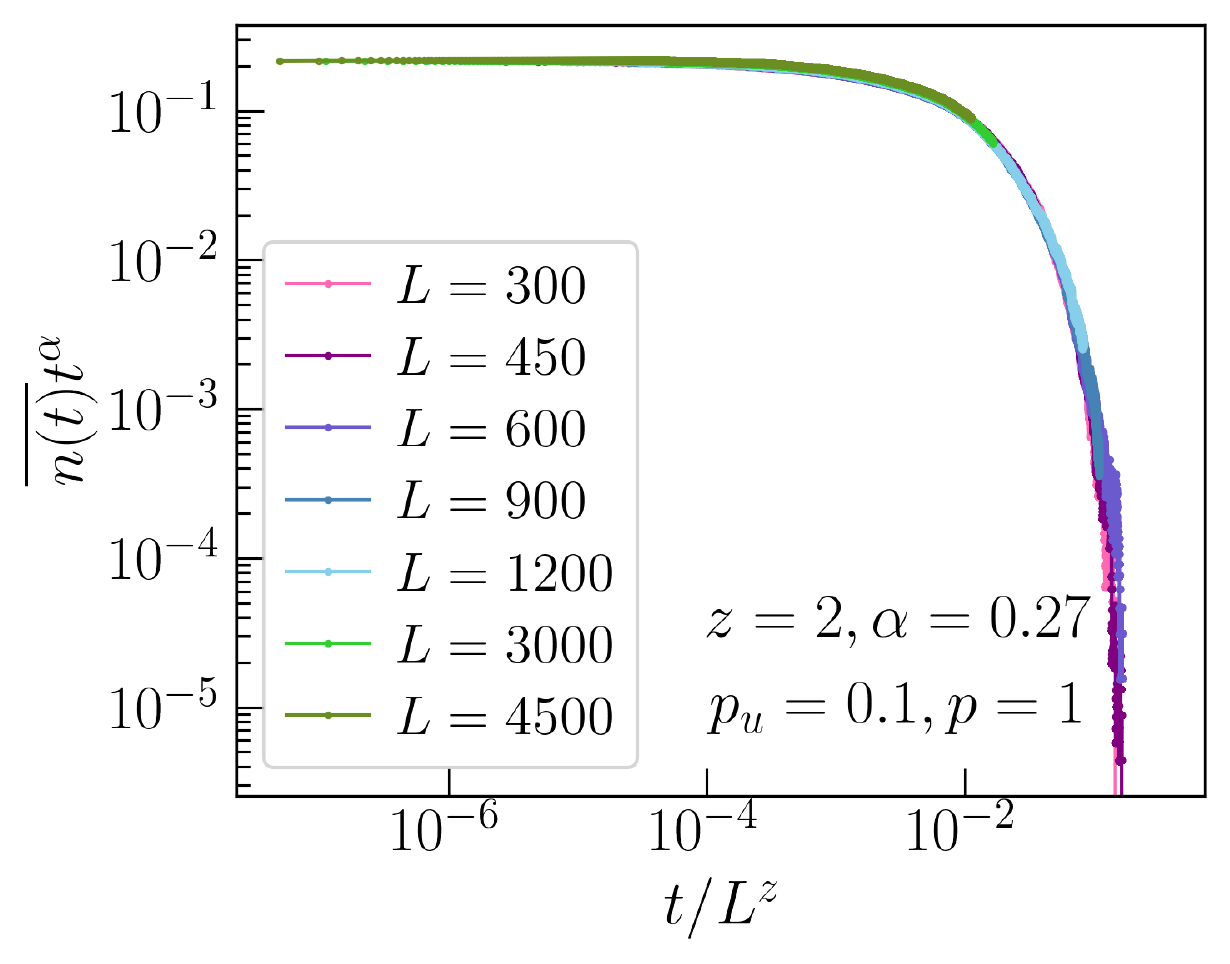}\label{fig:p_u=0.1,critical}}

  \subfigure[]{\includegraphics[width=0.4\textwidth]{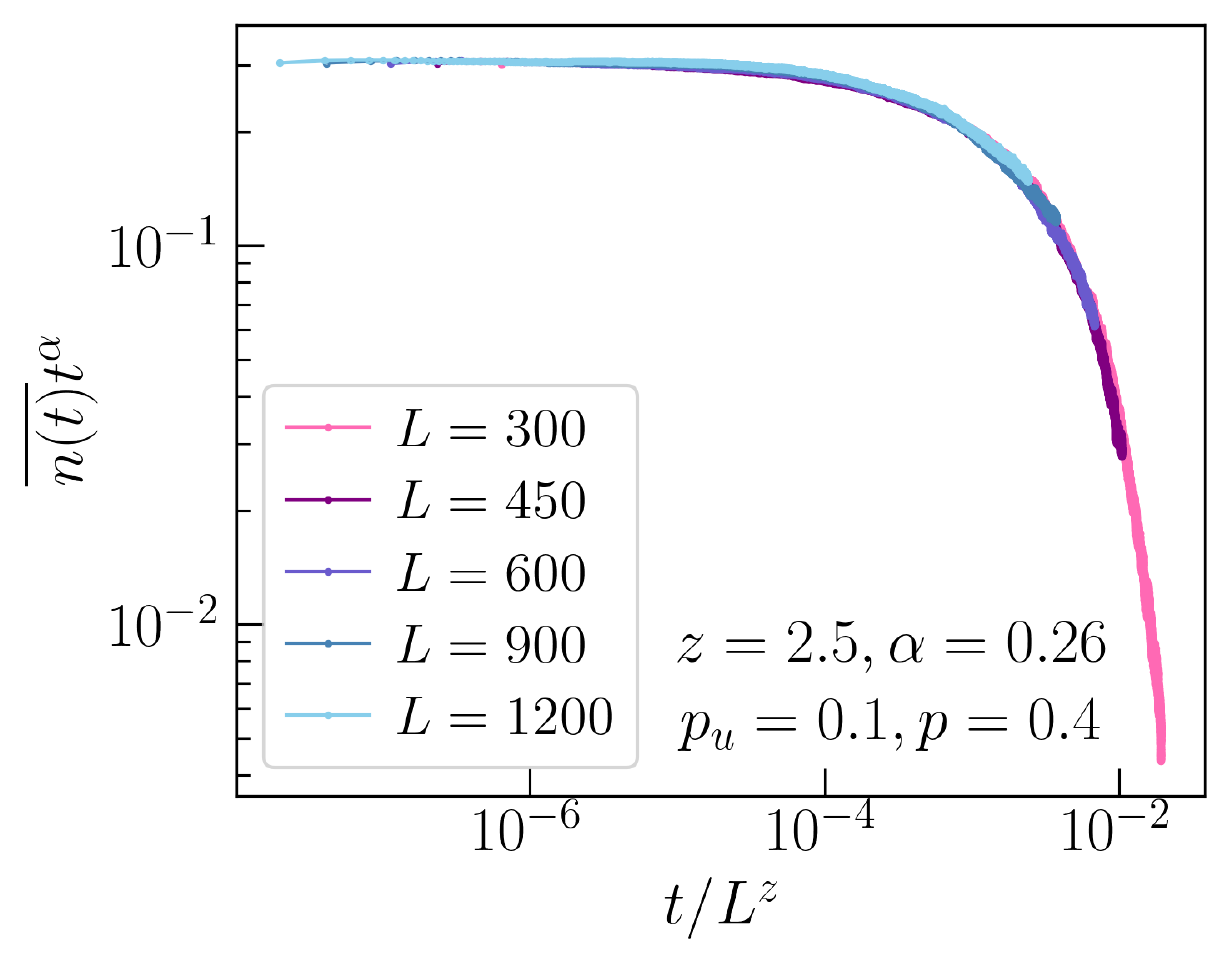}\label{fig:p_u=0.1,p_c}}%
  \subfigure[]{\includegraphics[width=0.4\textwidth]{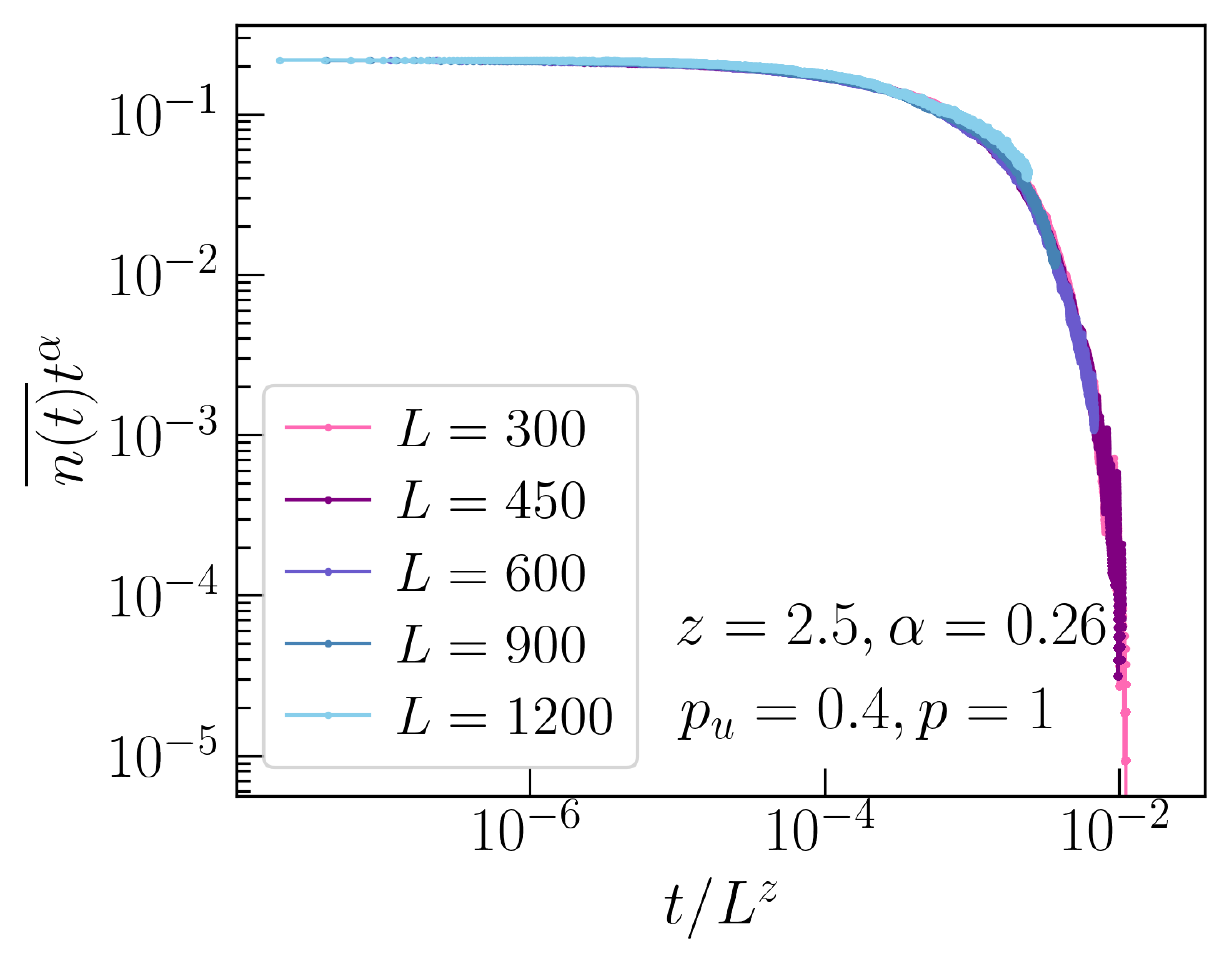}\label{fig:p_u=0.4,p_c}}
  \caption{The finite-size data collapse $\overline{n(t)}t^{\alpha}$ vs $t/L^z$ at: (a) $p_u=0$, $p=0.5$, with the dynamical exponent $z=1.95$ and $\alpha=0.28$. (b) $p_u=0.1$, $p=1$, with $z=2$ and $\alpha=0.27$. (c) $p_u=0.1$, $p=0.4$. (d) $p_u=0.4$, $p=1$. Both (c) and (d) have the same critical exponents $z=2.5$ and $\alpha=0.26$. The data are calculated over a variety of system sizes with the filling factor $\nu=1/3$ under PBC.}
  \label{fig:D_data_collapse}
\end{figure*}

Such a transition to a critical phase is a special feature of QA circuits and a similar transition has also been observed in the hybrid QA circuit with discrete $\mathbb{Z}_2$ symmetry \cite{QA_Z2}. To understand this phase transition, we can analyze the dynamics of the particles characterizing the bit string pairs difference, in particular, the particle density $n(t)\equiv\sum_x h(x,t)/L$ \cite{QA_Z2, QA_vl, Iaconis_2020}. In the $\mathbb{Z}_2$ symmetric QA circuit, it is shown that the particles perform branching-annihilating random walks (BAW) with an even number of offspring:
\begin{equation}
    W\leftrightarrow 3W, W+W\xrightarrow[]{p} \emptyset.
\end{equation}
The first process arises from unitary dynamics, while the second annihilation process occurs due to the measurement. The competition between these processes gives rise to a phase transition at a critical value $p_c$, which falls into the parity-conserving (PC) universality class. In the absorbing phase with $p>p_c$, the dynamics are primarily governed by the random walking particles, which  annihilate in pairs upon meeting. This particular dynamics leads to an algebraic decay with $n(t)\sim t^{-0.5}$, resulting in a power-law decay of $P(t)$. This, in turn, leads to a critical quantum phase characterized by logarithmic entanglement dynamics and a dynamical exponent of $z=2$.

In our model, the particle dynamics is similar and still preserves parity, but is more complicated.  For example, for the bit string pair $\{|011\rangle, |001\rangle\}$, under the Fredkin gate, it becomes $\{|110\rangle, |001\rangle\}$ and the particle representation $\circ\bullet\circ \mapsto \bullet\bullet\bullet$, i.e., the particles branch from 1 to 3. However, for the bit string pair $\{|111\rangle, |101\rangle\}$ which has the same particle representation as the previous one, it will remain invariant under the Fredkin gate. On the other hand, the SWAP gate enables particles to diffuse $\bullet\circ\leftrightarrow\circ\bullet$ regardless of the bit string configuration. Similarly, under the composite measurement, the particles will experience pair annihilation $\bullet\bullet\mapsto\circ\circ$ only when the bit string pair is $\{|10\rangle, |01\rangle\}$ instead of $\{|00\rangle,|11\rangle\}$. This leads to a phase transition belonging to a different universality class.

We first focus on the absorbing phase. In the limit $p_u=0$, we observe that the particle density follows a power law behavior $n(t)\sim t^{-\alpha}$ with $\alpha=0.28$  for all measurement rates $p>0$, which is significantly smaller than the exponent of $0.5$. This is because particle annihilation only occurs upon measurement based on specific bit string pairs, as illustrated earlier. More detailed data analysis suggests that $\overline{n(t)}$ for different system sizes can be collapsed onto the function
\begin{equation}
    \overline{n(t)}=t^{-\alpha}f(t/L^z),    
\end{equation}
where $z=1.95$, close to that of the critical phase observed in $\mathbb{Z}_2$ symmetric QA circuits. We also take a small but finite $p_u$ and similar scaling behavior is observed. For instance, in Fig.\ref{fig:p_u=0.1,critical}, we present the data collapse for $p_u=0.1$ and $p=1$, yielding exponents $\alpha=0.27$ and $z=2$. This absorbing phase with algebraic decay is responsible for a power law decay of $P(t)$, which further leads to a quantum critical phase with a logarithmic entanglement scaling. Notice that in this phase, $P(t)$  is mainly determined by the fraction $P^F(t)$ without extensively long domains. 

We further analyze the transition point. By fixing the unitary rate at $p_u=0.1$ and decreasing the measurement rate $p$, we can numerically identify the critical point. In Fig.\ref{fig:p_u=0.1,p_c}, it is observed that the critical point occurs at approximately $p=0.4$, with $\alpha=0.26$ and $z=2.5$. The existence of the phase transition persists as the unitary rate $p_u$ increases, until it reaches $p_u=0.4$. At this point (as illustrated in Fig.\ref{fig:p_u=0.4,p_c}), the critical point is found at the maximum allowed measurement rate $p=1$. It is worth emphasizing that this phase transition does not fall into the PC universality class, where the critical exponents are $z^{PC}=1.744$ and $\alpha^{PC}=0.286$.

\section{Discussions and outlook} 
In this paper, we investigate the entanglement dynamics of U(1) symmetric QA circuits. We show that the second R\'enyi entropy saturates the upper bound introduced in Eq.\eqref{eq:S_A upper bound}, namely $S_A^{(2)}(t)\propto\sqrt{t\ln{t}}$. To understand this behavior, we map the entanglement dynamics to a classical bit string model and demonstrate that the diffusive dynamics of $S_A^{(2)}(t)$ is caused by bit strings containing extensively long domains.

Additionally, we explore the monitored entanglement dynamics under U(1) symmetry and identify a phase transition from a volume-law phase to a critical phase as the measurement rate $p$ increases. Within the volume-law phase, $S_A^{(2)}(t)$ continues to exhibit diffusive growth due to the presence of bit strings with long domains that remain unaffected by the introduced measurements. On the other hand, the critical phase is characterized by logarithmic scaling of the entanglement, and its stability is ensured by both the U(1) symmetry and the basis-preserving nature of QA circuits.

The analysis of the second R\'enyi entropy can be extended to higher integer R\'enyi indices $n$. By making slight modifications to  Eq.\eqref{eq: purity_A}, it can be shown straightforwardly that the evolution of $S^{(n)}_A$ is mapped to a classical dynamics that encompasses n copies of bit strings. In particular, the volume-law phase exhibits diffusive dynamics in the presence of a logarithmic correction. This behavior is governed by these bit string configurations with extensive long domains. 

Notably, in both the volume-law phase and critical phase, the spin transport exhibits diffusive dynamics and fails to capture the entanglement phase transition. This is because this measurement-induced transition is visible solely in the non-linear observable of the density matrix. We confirm this diffusive transport by numerically computing the correlation functions and the detailed results are presented in Appendix.\ref{Appendix: Transport}. 

It has been established that the volume-law phase can be alternatively understood as a quantum error correcting code \cite{Gullans_2020, Choi_2020, Fan_2021, Li_2021, Li_2023}. Previously in Ref.~\cite{QA_vl}, we presented an interpretation of the quantum error correction property of the volume-law phase of a generic QA circuit, relating it to the dynamics of classical bit strings. Furthermore, we demonstrated a connection between quantum error-correcting codes and classical linear error-correcting codes.  In the case of the QA circuit with U(1) symmetry, we expect that the volume-law phase continues to exhibit the characteristics of a quantum error-correcting code. In particular, the dynamics of the associated classical bit strings reveal that the difference between bit string pairs still preserves classical information in a non-local manner, thereby functioning as a classical error-correcting code. It is worth noting that, unfortunately, this classical error-correcting code is no longer linear. We leave this for future study.

\acknowledgements
We thank Hisanori Oshima and Ethan Lake for the useful discussions. This research is supported in part by the Google Research Scholar Program and is supported in part by the National Science Foundation under Grant No. DMR-2219735. We gratefully acknowledge computing resources from Research Services at Boston College and the assistance provided by Wei Qiu. 

\appendix
\section{Two-species particle model}\label{Appendix: 2ps}

In Ref.~\cite{QA_Z2} and Ref.~\cite{QA_vl}, we proposed a two-species particle model which maps the entanglement dynamics of hybrid QA circuits under different symmetries to a classical two-species particle model. Before introducing the particle model, we will first give an overview of the classical bit string dynamics.

Since the charge-mixed state $|\psi_0\rangle=|+x\rangle^{L}$ has already been discussed in the main text, in this section we will focus on a charge-fixed initial state $|\psi_0\rangle=\frac{1}{\sqrt{N}}\sum_n |n\rangle$, where $\{|n\rangle=|\sigma_1\sigma_2\dots\sigma_L\rangle: \sigma_i=\{0,1\}, \sum_i\sigma_i=Q\}$ is the set of basis states with a fixed filling factor $\nu$, and $N={L \choose Q}=L!/[(L-Q)!Q!]$.

Then, the wavefunction at time $t$ can be represented as $|\psi_t\rangle=\tilde{U}_t|\psi_0\rangle$, where $\tilde{U}_t=M_t U_t M_{t-1} U_{t-1}\cdots$ represents the hybrid QA circuit of depth $t$ as an alternating combination of layers of measurements and unitary evolution. Recall that the second R\'enyi entropy $S_A^{(2)}=-\log_2{\text{Tr}(\rho_A^2)}$. The purity $\text{Tr}(\rho_A^2)$ equals the expectation value of the $\mathsf{SWAP}_A$ operator over two copies of the state \cite{SWAP, Islam_2015},
\begin{equation}
  \text{Tr}[\rho_A^2(t)]=\langle\psi_t|_2\otimes\langle\psi_t|_1 \mathsf{SWAP}_A|\psi_t\rangle_1\otimes|\psi_t\rangle_2.
\end{equation}
The wave function can be partitioned into subregions $A$ and $B$
\begin{equation}
  \begin{aligned}
    |\psi_t\rangle=\tilde{U}_t|\psi_0\rangle=\frac{1}{\sqrt{N}}\sum_{i,j}e^{i\theta_{ij}}|\alpha_i\rangle_A|\beta_j\rangle_{B},
  \end{aligned}
\end{equation}
The $\mathsf{SWAP}_A$ operator exchanges the spin configurations $|\alpha\rangle$ within subsystem $A$ of the double copies of $|\psi_t\rangle$. Then, we insert two sets of complete basis which we call ``bit strings'' \cite{Iaconis_2020},
\begin{widetext}
\begin{equation}\label{eq:purity_charge_fixed}
  \begin{aligned}
    \text{Tr}[\rho_A^2(t)]&=\sum_{n_1,n_2}\langle\psi_t|_2\langle\psi_t|_1 \mathsf{SWAP}_A|n_1\rangle|n_2\rangle\langle n_2|\langle n_1|\psi_t\rangle_1|\psi_t\rangle_2
    \\
    &=\sum_{n_1,n_2}\langle\psi_0|_1\tilde{U}_t^{\dagger}|n_1'\rangle\langle\psi_0|_2\tilde{U}_t^{\dagger}|n_2'\rangle \langle n_1|\tilde{U}_t|\psi_0\rangle_1\langle n_2|\tilde{U}_t|\psi_0\rangle_2
  \end{aligned}
\end{equation}
\end{widetext}
where 

\begin{equation}
\begin{aligned}
    |n_1'\rangle|n_2'\rangle&=\mathsf{SWAP}_A|n_1\rangle|n_2\rangle\\
&=\mathsf{SWAP}_A|\alpha_1\beta_1\rangle|\alpha_2\beta_2\rangle\\
&=|\alpha_2\beta_1\rangle|\alpha_1\beta_2\rangle.
\end{aligned}
\end{equation}
Since the dynamics preserves the total charge of the state, only the bit string configurations $\{|n_1\rangle,|n_2\rangle,|n_1'\rangle,|n_2'\rangle\}$ with the same filling factor $\nu$ will have non-zero overlap with $\langle\psi_0|$. We represent this subset of bit strings as  
\begin{widetext}
\begin{equation}
    \{n_1,n_2\}^{\nu}=\{|n_1\rangle=|\sigma^0_1\dots\sigma^0_L\rangle, |n_2\rangle=|\sigma^1_1\dots\sigma^1_L\rangle:\sum_{i=1}^L\sigma_i^\mu=\nu L, \sum_{i=1}^{L_A}\sigma_i^\mu+\sum_{i=L_A+1}^{L}\sigma_i^{1-\mu}=\nu L, \forall \mu\in\{0,1\}\}.
\end{equation}
\end{widetext}
Therefore, Eq.\eqref{eq:purity_charge_fixed} becomes
\begin{widetext}
\begin{equation}\label{eq:purity_final}
    \text{Tr}[\rho_A^2(t)]=\frac{1}{N^2}\sum_{\{n_1,n_2\}^{\nu}}e^{-i\Theta_{n_1'}(t)}e^{-i\Theta_{n_2'}(t)}e^{i\Theta_{n_1}(t)}e^{i\Theta_{n_2}(t)}.
\end{equation}
\end{widetext}

Strictly speaking, there does not exist $\tilde{U}_t^\dagger$ since the projective measurements are nonunitary operators. However, we can still deduce the effective action of the composite measurement $M_{1/2}$ on sites $i$ and $i+1$ of the bit string $|n\rangle\in\{n_1,n_2\}^\nu$, 
\begin{widetext}
\begin{equation}
    \langle n| M_{1/2}|\psi\rangle =
    \begin{cases}
    \langle n|\psi_0\rangle=\frac{1}{\sqrt{N}}e^{i\theta_n} & \text{if $\sigma_i=\sigma_{i+1}$} \\
    \langle T_{1/2}(n)|\psi_0\rangle = \frac{1}{\sqrt{N}}e^{i\theta_{T_{1/2}(n)}} & \text{if $\sigma_i\neq\sigma_{i+1}$}
    \end{cases}
\end{equation}
\end{widetext}
where $\langle T_{1/2}|$ stands for the bit string $\langle n|$ with the spins on sites $i$ and $i+1$ forced to be in the $|01\rangle/|10\rangle$ state. Hence, instead of following the quantum trajectory of $|\psi_t\rangle$, we can study the bit string dynamics in a time-reversed order, i.e., evaluate $\langle n|\tilde{U}_t|\psi_0\rangle$ from left to right,
\begin{equation}
  \begin{aligned}
    \langle n|\tilde{U}_t|\psi_0\rangle &=\langle n(t'=0)|M_t U_t M_{t-1} U_{t-1}\cdots|\psi_0\rangle
    \\
    &=\langle n(t'=1)|U_t M_{t-1} U_{t-1}\cdots|\psi_0\rangle
    \\
    &=e^{i\theta_{n(t'=1)}}\langle n(t'=1)|M_{t-1} U_{t-1}\cdots|\psi_0\rangle
    \\
    &\cdots
    \\
    &=\frac{1}{\sqrt{N}}e^{i\theta_{n(t'=1)}}e^{i\theta_{n(t'=2)}}\cdots e^{i\theta_{n(t'=t)}}
    \\
    &=\frac{1}{\sqrt{N}}e^{i\Theta_n(t)},
  \end{aligned}
\end{equation}
where $e^{i\Theta_n(t)}$ is one of the accumulated phase terms under time evolution that are multiplied and summed up over the ensemble of $\{|n_1\rangle,|n_2\rangle\}^\nu$ in Eq.~\eqref{eq:purity_final} to evaluate $\text{Tr}\rho_A^2$.

\begin{table*}[t]
\centering
\begin{tabular}{||c|c|c|c|c|c|c|c|c|c|c|c|c||}
    \hline
    \multirow{3}{4em}{Before (after) Fredkin} & $n_1$ & $|110\rangle$ & $|110\rangle$ & $|110\rangle$ & $|110\rangle$ & $|110\rangle$ & $|110\rangle$ & $|110\rangle$ & $|101\rangle$ & $|101\rangle$ & $|101\rangle$ & $|101\rangle$ \\
    & $n_2$ & $|101\rangle$ & $|011\rangle$ & $|100\rangle$ & $|010\rangle$ & $|001\rangle$ & $|111\rangle$ & $|000\rangle$ & $|100\rangle$ & $|010\rangle$ & $|001\rangle$ & $|111\rangle$ \\
    & $|n_1-n_2|$ & $\circ\bullet\bullet$ & $\bullet\circ\bullet$ & $\circ\bullet\circ$ & $\bullet\circ\circ$ & $\bullet\bullet\bullet$ & $\circ\circ\bullet$ & $\bullet\bullet\circ$ & $\circ\circ\bullet$ & $\bullet\bullet\bullet$ & $\bullet\circ\circ$ & $\circ\bullet\circ$ \\
    \hline
    \multirow{3}{4em}{After (before) Fredkin} & $n_1$ & $|011\rangle$ & $|011\rangle$ & $|011\rangle$ & $|011\rangle$ & $|011\rangle$ & $|011\rangle$ & $|011\rangle$  & $|101\rangle$ & $|101\rangle$ & $|101\rangle$ & $|101\rangle$ \\
    & $n_2$ & $|101\rangle$ & $|110\rangle$ & $|100\rangle$ & $|010\rangle$ & $|001\rangle$ & $|111\rangle$ & $|000\rangle$ & $|100\rangle$ & $|010\rangle$ & $|001\rangle$ & $|111\rangle$ \\
    & $|n_1-n_2|$ & $\bullet\bullet\circ$ & $\bullet\circ\bullet$ & $\bullet\bullet\bullet$ & $\circ\circ\bullet$ & $\circ\bullet\circ$ & $\bullet\circ\circ$ & $\circ\bullet\bullet$ & $\circ\circ\bullet$ & $\bullet\bullet\bullet$ & $\bullet\circ\circ$ & $\circ\bullet\circ$ \\
    \hline
    \hline
    \multirow{3}{4em}{Before (after) Fredkin} & $n_1$ & $|101\rangle$ & $|100\rangle$ & $|100\rangle$ & $|100\rangle$ & $|100\rangle$ & $|010\rangle$ & $|010\rangle$ & $|010\rangle$ & $|001\rangle$ & $|001\rangle$ & $|111\rangle$ \\
    & $n_2$ & $|000\rangle$ & $|010\rangle$ & $|001\rangle$ & $|111\rangle$ & $|000\rangle$ & $|001\rangle$ & $|111\rangle$ & $|000\rangle$ & $|111\rangle$ & $|000\rangle$ & $|000\rangle$\\
    & $|n_1-n_2|$ & $\bullet\circ\bullet$ & $\bullet\bullet\circ$ & $\bullet\circ\bullet$ & $\circ\bullet\bullet$ & $\bullet\circ\circ$ & $\circ\bullet\bullet$ & $\bullet\circ\bullet$ & $\circ\bullet\circ$ & $\bullet\bullet\circ$ & $\circ\circ\bullet$ & $\bullet\bullet\bullet$
    \\
    \hline
    \multirow{3}{4em}{After (before) Fredkin} & $n_1$ & $|101\rangle$ & $|100\rangle$ & $|100\rangle$ & $|100\rangle$ & $|100\rangle$ & $|010\rangle$ & $|010\rangle$ & $|010\rangle$ & $|001\rangle$ & $|001\rangle$ & $|111\rangle$ \\
    & $n_2$ & $|000\rangle$ & $|010\rangle$ & $|001\rangle$ & $|111\rangle$ & $|000\rangle$ & $|001\rangle$ & $|111\rangle$ & $|000\rangle$ & $|111\rangle$ & $|000\rangle$ & $|000\rangle$\\
    & $|n_1-n_2|$ & $\bullet\circ\bullet$ & $\bullet\bullet\circ$ & $\bullet\circ\bullet$ & $\circ\bullet\bullet$ & $\bullet\circ\circ$ & $\circ\bullet\bullet$ & $\bullet\circ\bullet$ & $\circ\bullet\circ$ & $\bullet\bullet\circ$ & $\circ\circ\bullet$ & $\bullet\bullet\bullet$
    \\
    \hline
\end{tabular}
\caption{The update rule of the particle configurations under the Fredkin gate $|\sigma_{i-1} 1_i \sigma_{i+1}\rangle \mapsto |\sigma_{i+1} 1_i \sigma_{i-1}\rangle$. The Fredkin gate only acts nontrivially on bit strings with $\sigma_i=1$ and $\sigma_{i-1}\neq\sigma_{i+1}$. For bit string pairs containing such configurations, the corresponding particles either branch/annihilate in pairs, i.e., $\bullet\bullet\bullet\leftrightarrow\circ\bullet\circ$, or diffuse i.e., $\circ\bullet\bullet\leftrightarrow\bullet\bullet\circ$, or $\bullet\circ\circ\leftrightarrow\circ\circ\bullet$.}
\label{table:Fredkin update rule}
\end{table*}

\begin{table*}
\centering
\begin{tabular}{||c|c|c|c|c|c|c|c||}
    \hline
    \multirow{3}{3em}{Before $M_{1/2}$} & $n_1$ & $|10\rangle$ & $|10\rangle$ & $|10\rangle$ & $|01\rangle$ & $|01\rangle$ & $|11\rangle$
    \\
    & $n_2$ & $|01\rangle$ & $|11\rangle$ & $|00\rangle$ & $|11\rangle$ & $|00\rangle$ & $|00\rangle$ \\
    & $|n_1-n_2|$ & $\bullet\bullet$ & $\circ\bullet$ & $\bullet\circ$ & $\bullet\circ$ & $\circ\bullet$ & $\bullet\bullet$ \\
    \hline    
    \multirow{3}{3em}{After $M_1$} & $n_1$ & $|01\rangle$ & $|01\rangle$ & $|01\rangle$ & $|01\rangle$ & $|01\rangle$ & $|11\rangle$ \\
    & $n_2$ & $|01\rangle$ & $|11\rangle$ & $|00\rangle$ & $|11\rangle$ & $|00\rangle$ & $|00\rangle$ \\
    & $|n_1-n_2|$ & $\circ\circ$ & $\bullet\circ$ & $\circ\bullet$ & $\bullet\circ$ & $\circ\bullet$ & $\bullet\bullet$ \\
    \hline
    \multirow{3}{3em}{After $M_2$} & $n_1$ & $|10\rangle$ & $|10\rangle$ & $|10\rangle$ & $|10\rangle$ & $|10\rangle$ & $|11\rangle$ \\
    & $n_2$ & $|10\rangle$ & $|11\rangle$ & $|00\rangle$ & $|11\rangle$ & $|00\rangle$ & $|00\rangle$ \\
    & $|n_1-n_2|$ & $\circ\circ$ & $\circ\bullet$ & $\bullet\circ$ & $\circ\bullet$ & $\bullet\circ$ & $\bullet\bullet$ \\
    \hline
\end{tabular}
\caption{The update rule of the particle configurations under the composite measurement $M_{1/2}=R\circ P_{1/2}$. Note that the composite measurement only acts nontrivially on bit string configurations with anti-parallel neighboring spins. For bit string pairs containing such configurations, under the measurement, the corresponding particle representation either annihilates in pairs, i.e., $\bullet\bullet \mapsto \circ\circ$, or diffuses, i.e., $\bullet\circ\leftrightarrow\circ\bullet$.}
\label{table:Measurement update rule}
\end{table*}

In order to understand the dynamics of the relative phase $\Theta_r=-\Theta_{n_1'}-\Theta_{n_2'}+\Theta_{n_1}+\Theta_{n_2}$, we consider the evolution of the difference between a bit string pair $\{|n_1\rangle,|n_2\rangle\}$,
\begin{equation}
  h(x,t)=|n_1(x,t)-n_2(x,t)|.
\end{equation}
It is then natural to use the particle representation where the empty site symbol $\circ$ denotes $h(x)=0$ and the occupied site symbol $\bullet$ denotes $h(x)=1$. Specifically, we represent the difference at $t=0$ in subregion $A$ (subregion $B$) by $X$ ($Y$) particles. Let $\tilde{X}/\tilde{Y}$ denote the rightmost $X$ (leftmost $Y$) particle and $x/y$ denote their positions. As shown in Fig. \ref{fig:two_species}, under the time evolution, the particles start to evolve according to the update rule. As mentioned in the main text, to determine the update rule of the particles under the U(1)-symmetric circuit, one should always refer to the corresponding bit string dynamics. The detailed update rule is listed in Table.\ref{table:Fredkin update rule} and Table.\ref{table:Measurement update rule}. Before the two species encounter each other, the phase generated by each layer of unitary evolution on $|n\rangle$ is $\theta_n=\theta_n^{[1,x]}+\theta_n^{(x,y)}+\theta_n^{[y,L]}$, i.e., the sum of phases generated within the regimes $[1,x]$, $(x,y)$ and $[y,L]$. The bit string configurations within $[1,x]$ occupied by $X$ particles always satisfy $n_1([1,x])=n_2'([1,x])$ and $n_2([1,x])=n_1'([1,x])$. Therefore, $\theta_{n_1}^{[1,x]}=\theta_{n_2^\prime}^{[1,x]}$ and $\theta_{n_2}^{[1,x]}=\theta_{n_1^\prime}^{[1,x]}$. Similarly, for the regime occupied by $Y$ particles, since $n_1([y,L])=n_1'([y,L])$ and $n_2([y,L])=n_2'([y,L])$, we always have $\theta_{n_1}^{[y,L]}=\theta_{n_1^\prime}^{[y,L]}$ and $\theta_{n_2}^{[y,L]}=\theta_{n_2^\prime}^{[y,L]}$. At the same time, since there is no bit string difference within the regime $(x,y)$, $\theta_{n_1}^{(x,y)}=\theta_{n_2}^{(x,y)}=\theta_{n_1^\prime}^{(x,y)}=\theta_{n_2^\prime}^{(x,y)}$. Therefore, the phase difference along the lattice vanishes: $-\theta_{n_1'}-\theta_{n_2'}+\theta_{n_1}+\theta_{n_2}=0$. If for a bit string pair $\{|n_1\rangle,|n_2\rangle\}$, $X$ and $Y$ particles do not meet each other up to time $t$, then the accumulated relative phase $\Theta_{r}(t)$ is zero and such pair contributes $1/4^L$ to the purity $\text{Tr}[\rho_A^2(t)]$.

Once the rightmost $X$ particle comes across the leftmost $Y$ particle, the two-qubit phase gate acting on sites $x$ and $y$ will generate a nonzero relative phase. For example, if we apply the CZ gate on $\bullet\red{\bullet}$ with a possible corresponding bit string configuration $\{|n_1\rangle,|n_2\rangle,|n_1'\rangle,|n_2'\rangle\}_{x,y}=\{|10\rangle,|01\rangle,|00\rangle,|11\rangle\}$, a relative phase $0+0-0-\pi=-\pi$ is generated. If we apply the CNOT gate on sites $x$ and $y$, $\{|n_1\rangle,|n_2\rangle,|n_1'\rangle,|n_2'\rangle\}_{x,y}\to\{|11\rangle,|01\rangle,|00\rangle,|10\rangle\}$, i.e., another type of ``particle'' different from the two species with bit string configuration $|n_1\rangle_{y}=|n_2\rangle_{y}\neq|n_1'\rangle_{y}=|n_2'\rangle_{y}$ appears on site $y$ and will spread along the lattice under further evolution. As time evolves, the configurations for which the two species have met will generate random accumulated phases, half of which are composed of odd numbers of $\pi$, while the other half are composed of even numbers of $\pi$. The accumulated phase terms $e^{i\Theta_r}$ of such configurations will add up to zero and make no contribution to Eq.~\eqref{eq:purity_final}. Therefore, we have 
\begin{equation}
  \begin{aligned}
    &\text{Tr}\rho_A^2(t)\approx P(t),
    \\
    &S_A^{(2)}(t)\approx -\ln{P(t)},
  \end{aligned}
  \end{equation}
where $P(t)$ is the fraction of particle configurations in which $X$ and $Y$ particles never encounter one another up to time $t$. This quantum-classical correspondence has been numerically verified in Ref.~\cite{QA_Z2}.

\section{Entanglement dynamics of U(1) symmetric QA circuits with other kinetic constraints}\label{Appendix: other models}
\subsection{The SWAP Model}

\begin{figure}[!t]
  \centering
  \includegraphics[width=0.45\textwidth]{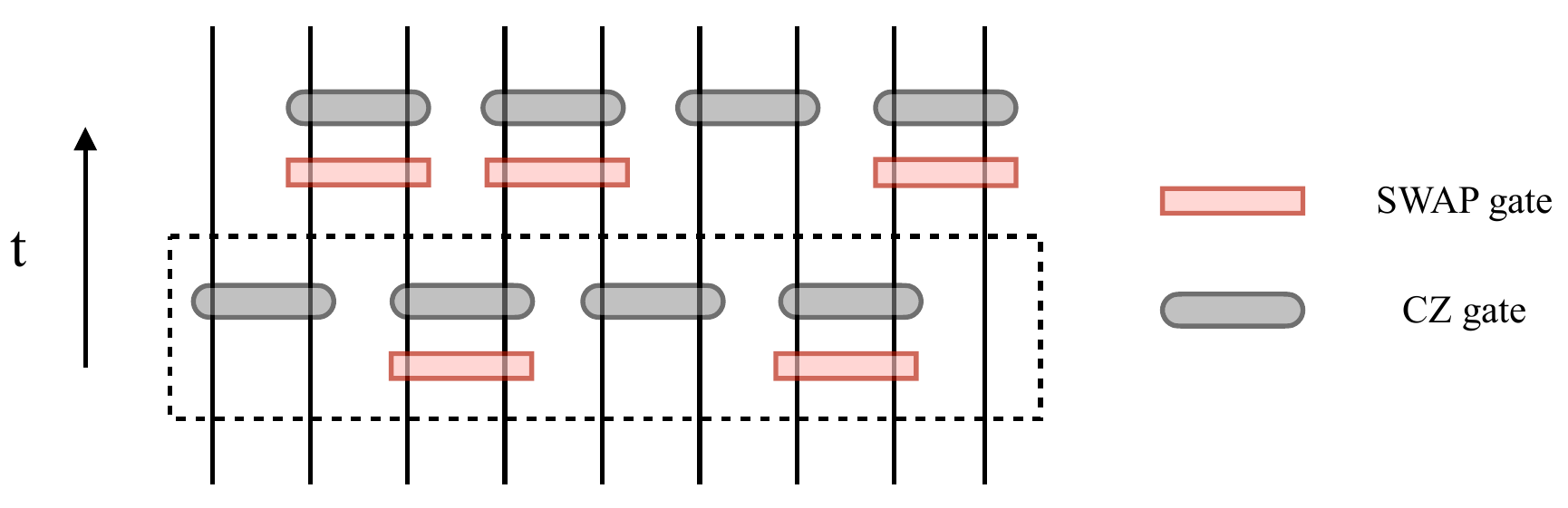}
  \caption{The arrangement of gates of the SWAP model in a time step. The dashed box encloses the gates within a single layer. Each time step involves two layers of SWAP gates with probability $p_u=0.5$, and CZ gates.}
  \label{fig:SWAP_illustration}
\end{figure}

\begin{figure}
  \centering
  \subfigure[]{\includegraphics[width=0.4\textwidth]{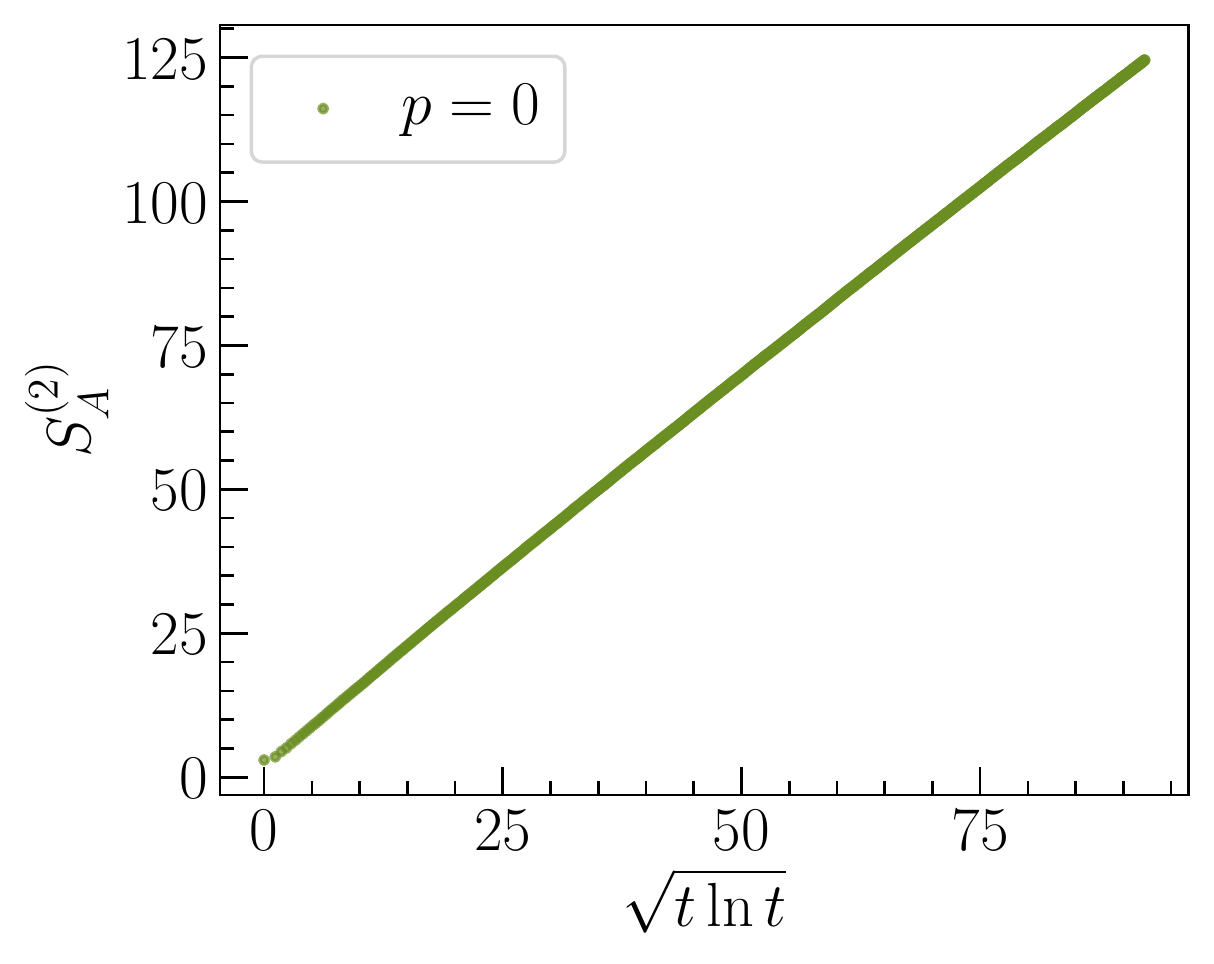}\label{fig:SWAP_SA}}
  \subfigure[]{\includegraphics[width=0.4\textwidth]{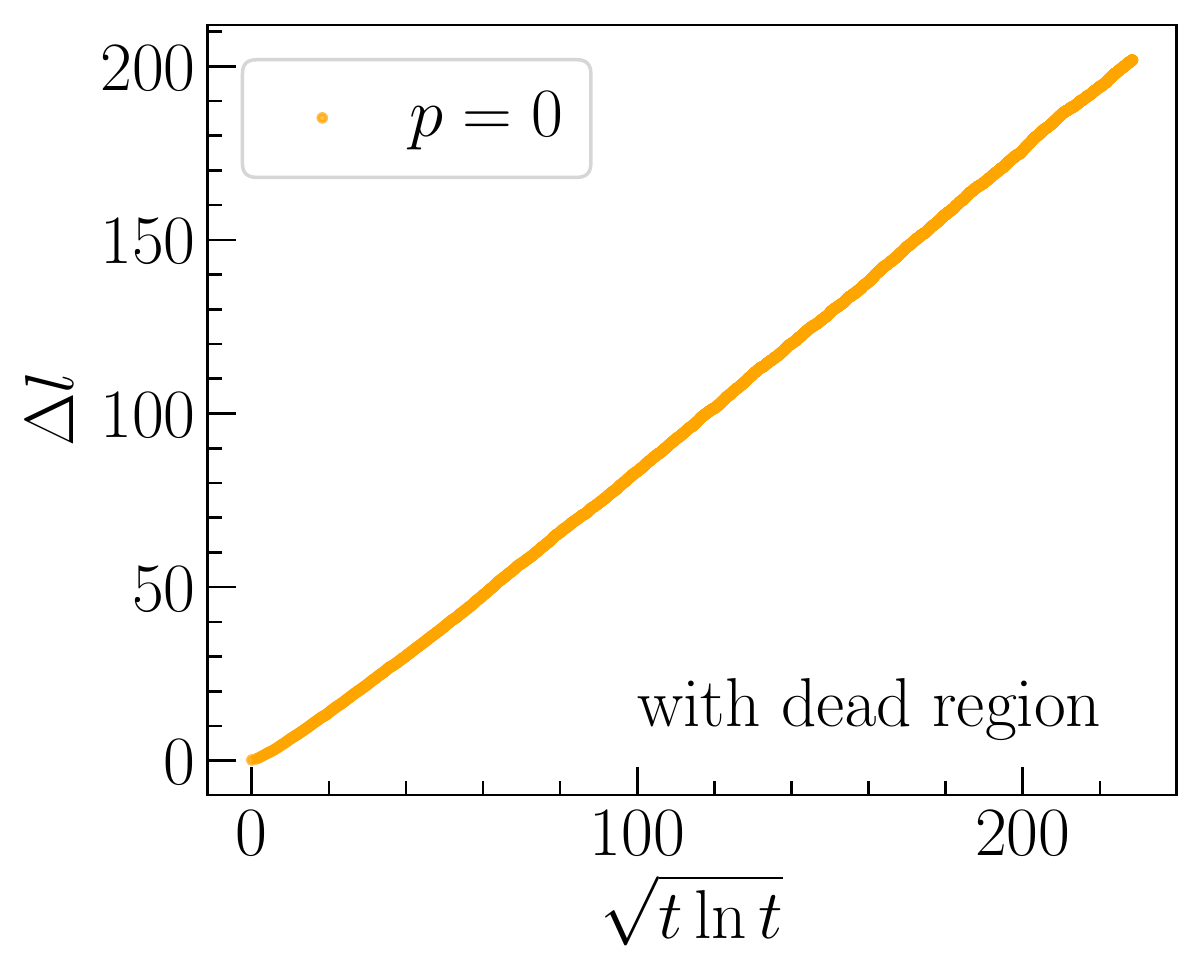}\label{fig:SWAP_l}}
  \caption{(a) $\overline{S_A^{(2)}(t)}$ of the SWAP model simulated using the stabilizer formalism. (b) The $\tilde{X}$ particle displacement $\overline{\Delta l(t)}$ of the slow modes. It is found that both $\overline{S_A^{(2)}(t)}$ and $\overline{\Delta l(t)}$ scale as $\sqrt{t\ln{t}}$. We take the system size $L=600$ and $|\psi_0\rangle=|+x\rangle^{\otimes L}$ for (a) and the bit strings with dead regions with the charge filling $\nu_A=1/3$ in subsystem $A$ for (b). }
\end{figure}

We study the entanglement dynamics under the kinetic constraint determined by SWAP gates solely, so that our circuit is a U(1) symmetric Clifford QA circuit, which can be efficiently simulated at large system sizes using stabilizer formalism \cite{Stabilizer}. We will only consider the unitary dynamics as the composite measurements can not be simulated by stabilizer formalism. As illustrated in Fig.\ref{fig:SWAP_illustration}, we consider a circuit where each time step involves two layers of SWAP gates and CZ gates, with each applying on odd/even sites. To achieve enough randomness, we take the SWAP rate to be $p_u=0.5<1$. In addition, we consider $|\psi_0\rangle=|+x\rangle^{\otimes L}$ as the initial state and take $L=600$. As shown in Fig.\ref{fig:SWAP_SA}, the entanglement entropy grows diffusively with a logarithmic correction, i.e., $\overline{S_A^{(2)}}\propto \sqrt{t\ln{t}}$. 

On the other hand, we examine the endpoint displacement $\overline{\Delta l(t)}$ of the bit strings with the dead region with a charge filling factor of $\nu_A=1/3$. The numerics in Fig.\ref{fig:SWAP_l} indicates that $\overline{\Delta l(t)}\propto\sqrt{t\ln{t}}$ as well. This scaling can be explained by exactly mapping the spin dynamics of the SWAP model to the simple symmetric exclusion processes. In this process, each charge undertakes a symmetric random walk, while being prohibited from jumping to an already occupied site. It has been analytically shown in Ref.~\cite{symmetric_exclusion} that the position of the rightmost charge expands as $\sqrt{t\ln t}$.

It is worth noting that Clifford circuits with U(1) symmetry are highly restricted. For example, all R\'enyi entropies are equal for Clifford circuits, which fails to capture the ballistic growth of von Neumann entropy for generic U(1) symmetric random circuits. Nevertheless, the SWAP model enables us to verify at a large system size that the second R\'enyi entropy indeed saturates the upper bound in Eq.~\eqref{eq:S_A upper bound} and is dominated by rare slow modes with extensively long domains.

\subsection{The Four-qubit CSWAP model}
\begin{figure}[!t]
  \centering
  \includegraphics[width=0.45\textwidth]{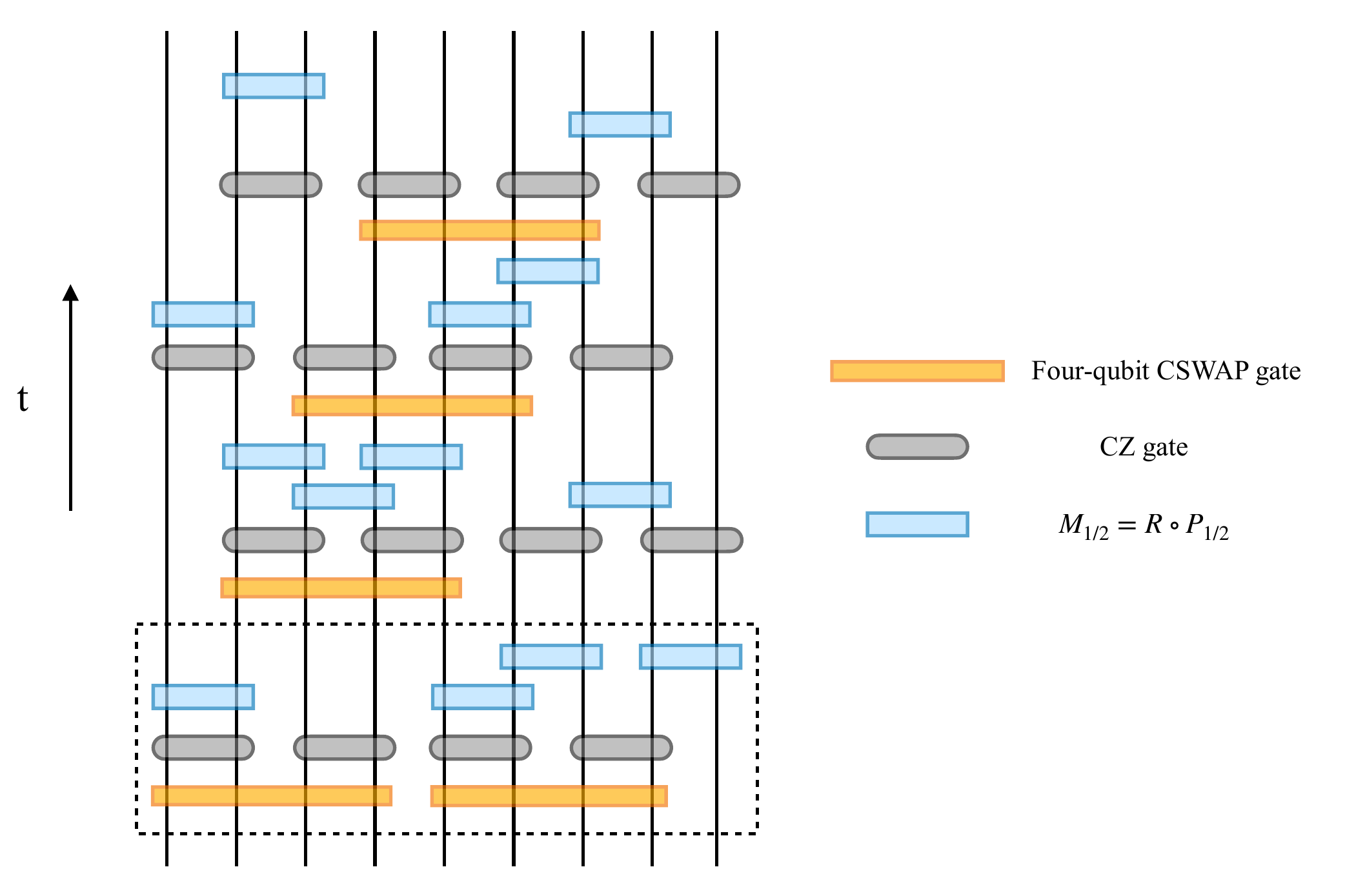}
  \caption{The arrangement of gates of the Four-qubit CSWAP model in a time step. The dashed box encloses the gates within a single layer. Each time step involves four layers of the four-qubit CSWAP gates with probability $p_u$, and CZ gates, interspersed with measurements with probability $p$.}
  \label{fig:CSWAP_illustration}
\end{figure}

\begin{figure}
  \centering
  \subfigure[]{\includegraphics[width=0.4\textwidth]{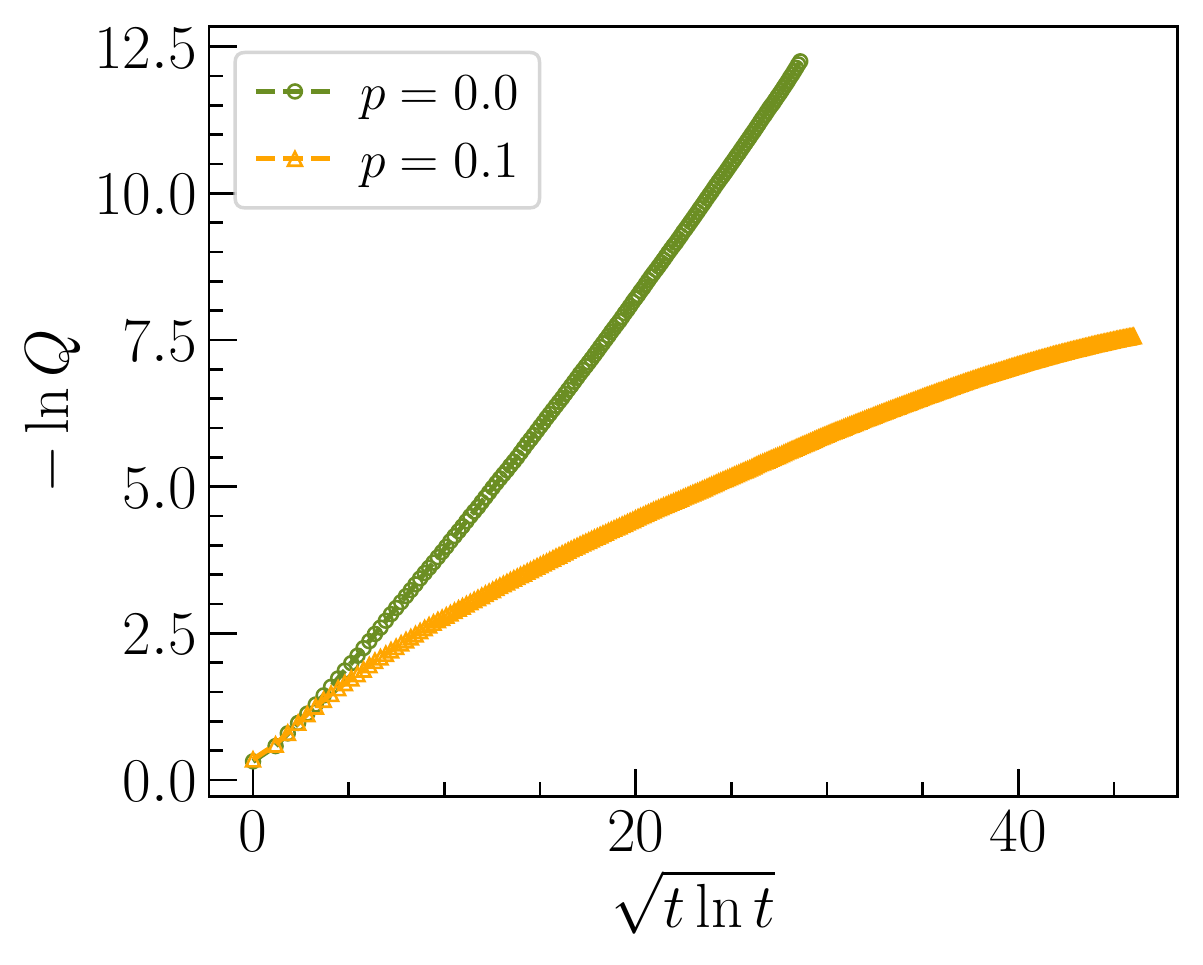}}
  \subfigure[]{\includegraphics[width=0.4\textwidth]{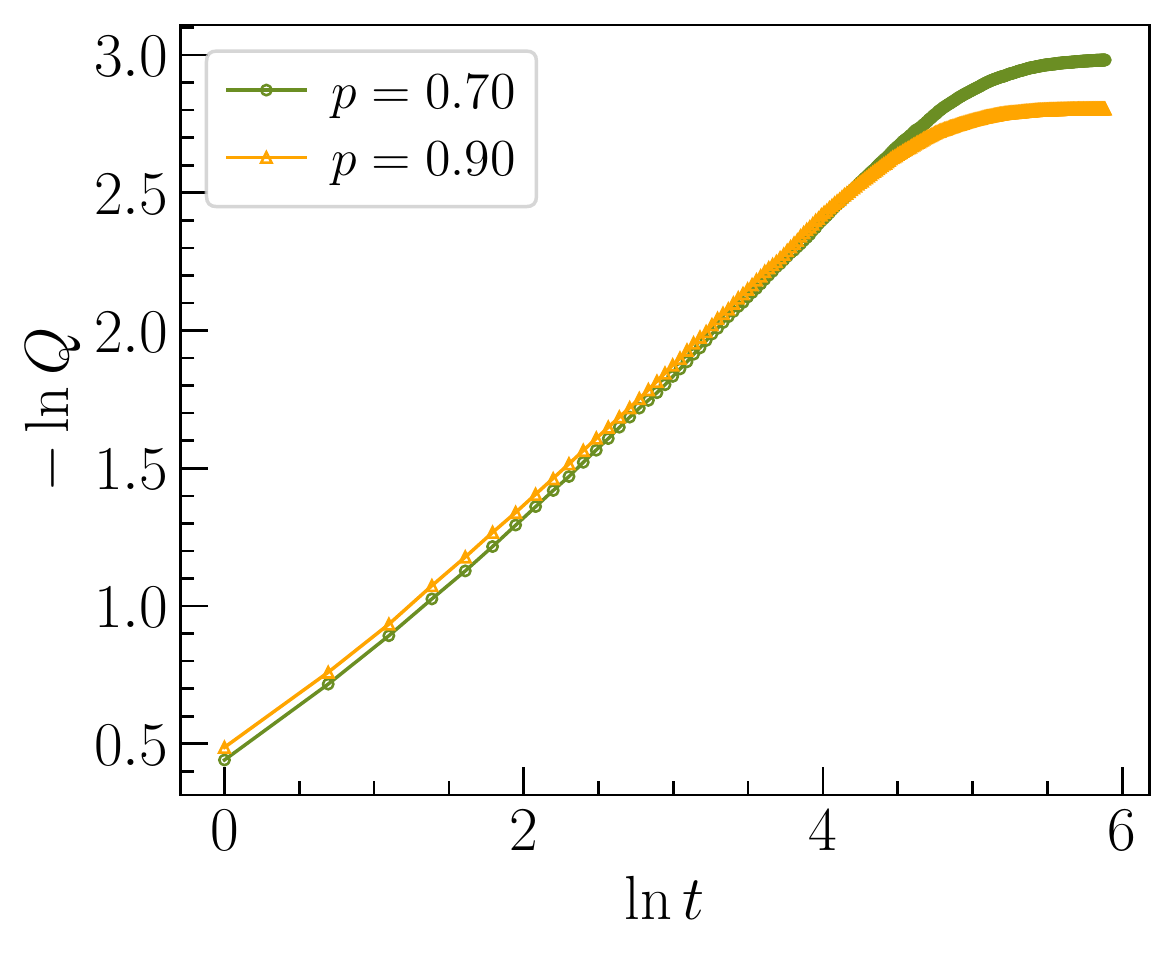}}
  \caption{$-\overline{\ln{Q}}$ of the Four-qubit CSWAP model at the measurement rate (a) $p=0$ and $p=0.1$, and (b) $p=0.7$ and $p=0.9$. We observe that $-\overline{\ln Q}\propto \sqrt{t\ln{t}}$ for (a) and $-\overline{\ln Q}\propto \ln{t}$ for (b). We take the system size $L=120$, $p_u=0.5$, and charge filling $\nu=1/2$.}
  \label{fig:CSWAP}
\end{figure}

\begin{figure}
  \centering
  \includegraphics[width=0.4\textwidth]{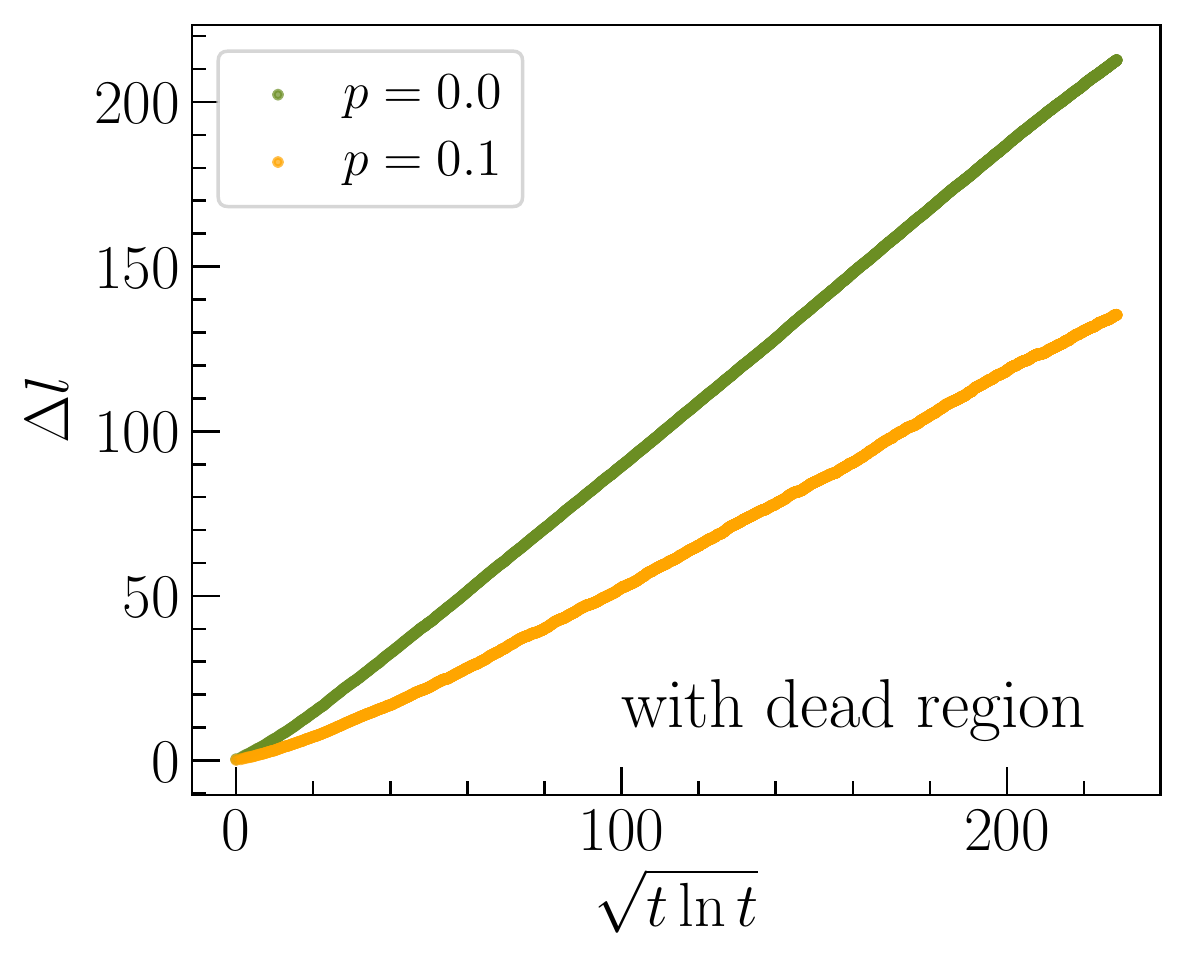}
  \caption{The endpoint displacement $\overline{\Delta l(t)}$ of the bit strings with the dead region of the Four-qubit CSWAP model vs $\sqrt{t\ln{t}}$ at $p=0$ and $p=0.1$. We take the system size $L=600$, $p_u=0.5$, and $\nu_A=1/2$.}
  \label{fig:CSWAP_l}
\end{figure}

Now we consider the entanglement dynamics of the U(1) symmetric QA circuits with the kinetic constraint determined by a four-qubit gate, which swaps the spins $\sigma_2$ and $\sigma_3$ if the first spin $\sigma_1=0$ or the fourth spin $\sigma_4=1$. It is also called Fredkin gate in other works \cite{Yang_2022, Singh_2021}, to distinguish it from the Fredkin gate in the main text, we call it the Four-qubit CSWAP gate. As shown in Fig.\ref{fig:CSWAP_illustration}, each time step of the circuit consists of four layers of gates under PBC and in each layer, the Four-qubit CSWAP gates are applied on sites $\{4j-3, 4j-2, 4j-1, 4j\}/\{4j-2, 4j-1, 4j, 4j+1\}/\{4j-1, 4j, 4j+1, 4j+2\}/\{4j, 4j+1, 4j+2, 4j+3\}$ for $j\in[1, L/4]$ with probability $p_u=0.5$, and the CZ gates are applied on sites $\{2j-1,2j\}/\{2j,2j+1\}/\{2j-1,2j\}/\{2j,2j+1\}$, interspersed with composite measurements with probability $p$ applied on both odd and even sites.

To simplify the numerical simulation, we can fix the position of $\tilde{Y}$ particle to be the boundary between subsystems $A$ and $B$. This is equivalent to focusing on the single species picture where there are only $X$ particles and considering a subset of the phase difference in Eq.~\eqref{eq:purity_final}, i.e., the phase difference of $|n_1\rangle$ and $|n_1'\rangle$ restricted in regime $B$, denoted as
\begin{equation}
    Q\equiv\frac{1}{M}\sum_{n_1,n_1'}e^{-i\Theta_{n_1'}^B+i\Theta_{n_1}^B},
\end{equation}
where $M$ is the number of bit string pairs $\{|n_1\rangle,|n_1'\rangle\}$. With this approximation, the configurations which contribute to $Q$ are those whose $\tilde{X}$ particles never reach the middle cut.

We numerically simulate $-\ln{Q(t)}$ for an ensemble of bit strings $\{|n_1\rangle, |n_1'\rangle\}$ with system size $L=120$ and charge filling $\nu=1/2$. As shown in Fig.\ref{fig:CSWAP}, there exists a similar phase transition from a volume-law phase to a critical phase: when $p<p_c$, $-\overline{\ln Q}\propto \sqrt{t\ln{t}}$, and when $p>p_c$, $-\overline{\ln Q}\propto\ln{t}$. We believe that this applies to $\overline{S_A^{(2)}(t)}$ as well.

Finally, we study the endpoint displacement $\Delta l(t)$ of the bit strings with the dead region evolved under the Four-qubit CSWAP gates. The numerics in Fig.\ref{fig:CSWAP_l} confirms that $\overline{\Delta l}\propto \sqrt{t\ln{t}}$ for pure unitary dynamics and small measurement rate $p=0.1$. We believe that this diffusive growth persists in the whole volume-law phase.

\section{Spin transport of hybrid QA circuits with U(1) symmetry}\label{Appendix: Transport}

\begin{figure}[!t]
  \centering
  \subfigure[]{\includegraphics[width=0.4\textwidth]{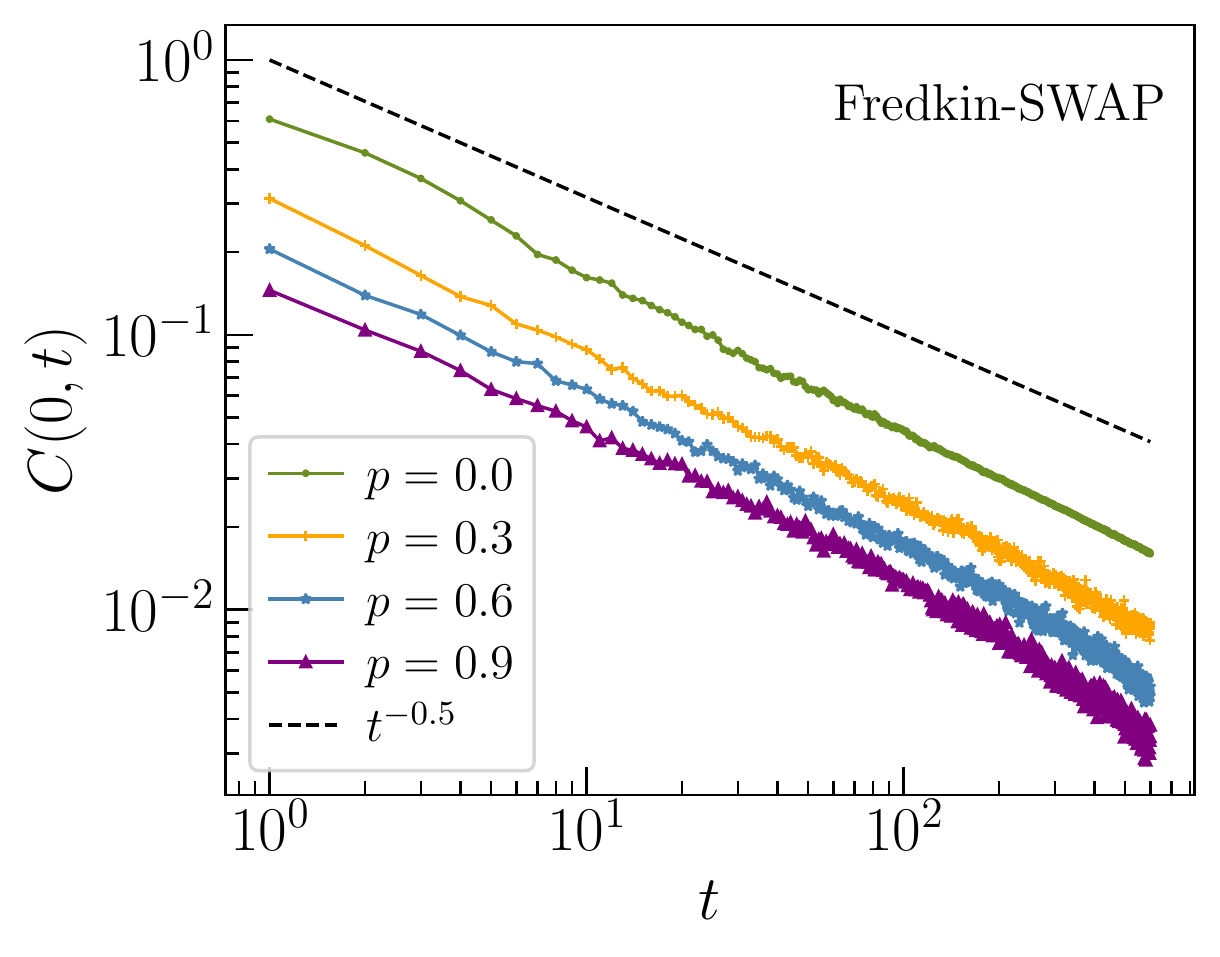}\label{fig:C_Fredkin_SWAP}}
  \subfigure[]{\includegraphics[width=0.4\textwidth]{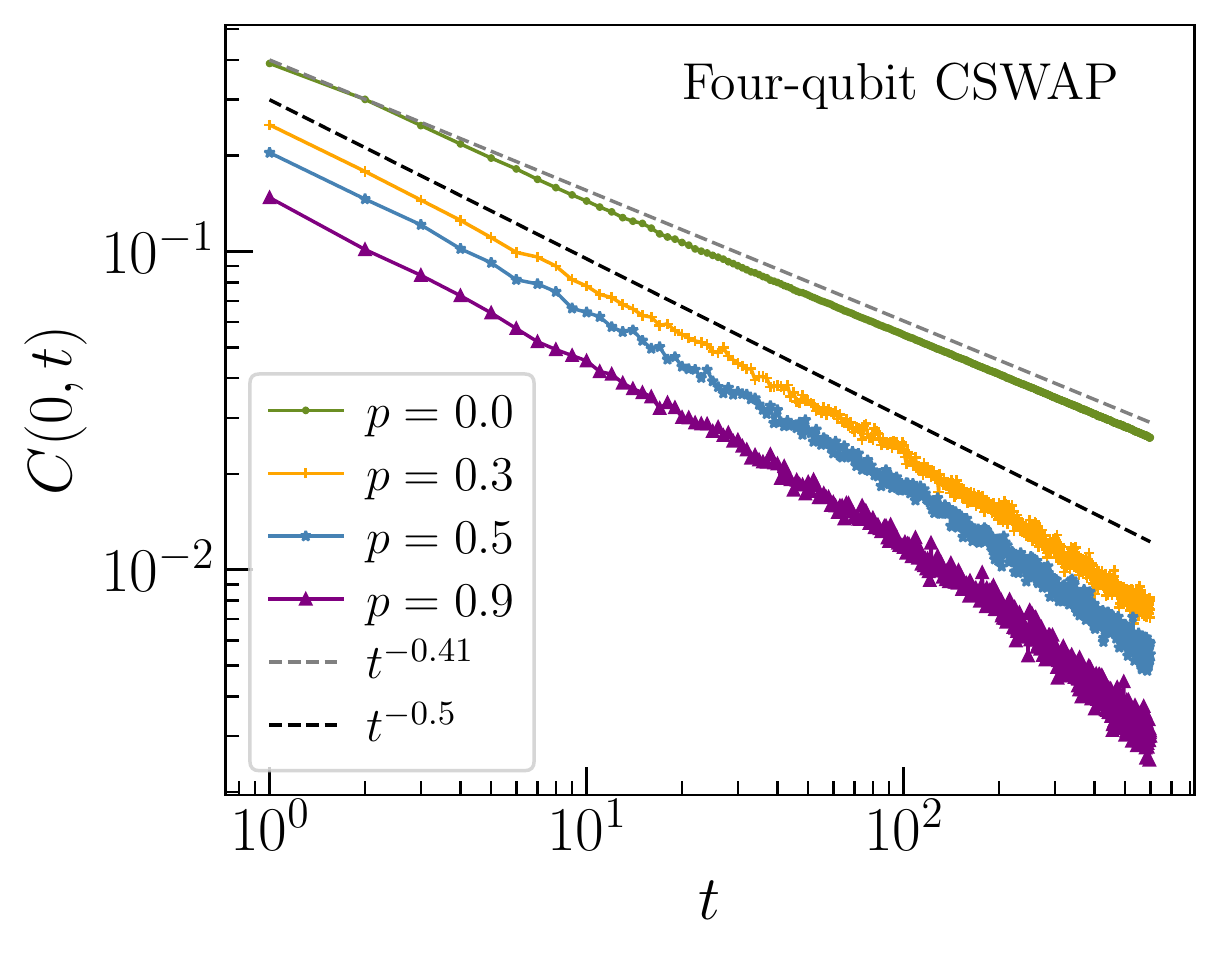}\label{fig:C_4CSWAP}}
  \caption{The correlation function $C(0,t)$ of (a) the Fredkin-SWAP model and (b) the Four-qubit CSWAP model. We take the system size $L=300$ and $p_u=0.1$, $\nu=1/3$ for (a), and $p_u=0.5$, $\nu=1/2$ for (b). It is found that for (a), $C(0,t)\propto t^{-0.5}$ for all $p$, and for (b), $C(0,t)\propto t^{-0.41}$ for $p=0$ and $C(0,t)\propto t^{-0.5}$ for $p>0$.}
\end{figure}

In this section, we consider the transport properties of the conserved charges which can be characterized by the spin correlation function
\begin{equation}
    C(x,t)=\langle Z_x(t)Z_0(0) \rangle - \langle Z_x(t)\rangle\langle Z_0(0)\rangle,
\end{equation}
where site 0 is in the middle of the system. If we consider a charge-fixed state $|\psi_0\rangle=\frac{1}{\sqrt{N}}\sum_{n}|n\rangle$ with the filling factor $\nu$, then the correlator of our QA circuit can be sampled using the classical bit strings via
\begin{widetext}
\begin{equation}
\begin{aligned}
    C(x,t)&=\frac{1}{N}\sum_{n}\langle n|\tilde{U}_t^\dagger Z_x(0)\tilde{U}_tZ_0(0)|n\rangle -\frac{1}{N}\sum_{n}\langle n|\tilde{U}_t^\dagger Z_x(0)\tilde{U}_t|n\rangle\frac{1}{N}\sum_{m}\langle m|Z_0(0)|m\rangle \\
    &=\frac{1}{N}\sum_n Z_{n(t),x} Z_{n,0}-\frac{1}{N^2}\sum_n Z_{n(t),x}\sum_{m}Z_{m,0},
\end{aligned}
\end{equation}
\end{widetext}
where $Z_{n(t),x}$ is the spin value at site $x$ of the bit string $|n(t)\rangle$ at time $t$. The correlation function for different system sizes can be collapsed onto the scaling form
\begin{equation}
    C(x,t)=t^{-1/z}f(x/t^{1/z}).
\end{equation}
Here we will only focus on the correlation in the time direction $C(0,t)$ for the Fredkin-SWAP model and the Four-qubit CSWAP model, and observe the dynamical exponents as we vary the measurement rate $p$.

As shown in Fig.\ref{fig:C_Fredkin_SWAP}, for the Fredkin-SWAP model, $C(0,t)\propto t^{-1/z}$ with $z=2$ for all $p$. The spin transport is diffusive with and without the measurements and hence fails to reflect the measurement-induced entanglement phase transition. This is because MIPTs are only visible in observables that are nonlinear in the density matrix. Similar diffusive dynamics is observed for all $p>0$ in the Four-qubit CSWAP model, as shown in Fig.\ref{fig:C_4CSWAP}. Interestingly, when $p=0$, the dynamical exponent $z=2.44>2$, consistent with $z=8/3$ as discovered in previous literature \cite{Yang_2022,Singh_2021}. Nevertheless, the spin at the boundary of the extensively long domain still exhibits diffusive dynamics with a logarithmic correction.

\bibliographystyle{quantum}
\bibliography{reference}
\end{document}